\documentclass[graybox, nosecnum]{svmult}

\usepackage{amssymb}		
\usepackage{amsmath}

\usepackage{mathptmx}       
\usepackage{helvet}         
\usepackage{courier}        
\usepackage{type1cm}        
\usepackage{makeidx}         
\usepackage{graphicx}        
\usepackage{multicol}        
\usepackage[bottom]{footmisc}
\usepackage{hyperref}        
\usepackage{soul}            

\usepackage[numbers,sort&compress]{natbib}

\makeindex             

\begin{document}
\title*{Influence of nuclear structure in relativistic heavy-ion collisions}
\author{Yu-Gang Ma \thanks{corresponding author} and Song Zhang}
\institute{Yu-Gang Ma \at Key Laboratory of Nuclear Physics and Ion-beam Application (MOE), Institute of Modern Physics, Fudan University, Shanghai 200433, China;\\ Shanghai Research Center for Theoretical Nuclear Physics, NSFC and Fudan University, Shanghai 200438, China\\ \email{mayugang@fudan.edu.cn}
\and Song Zhang \at Key Laboratory of Nuclear Physics and Ion-beam Application (MOE), Institute of Modern Physics, Fudan University, Shanghai  200433, China;\\ Shanghai Research Center for Theoretical Nuclear Physics, NSFC and Fudan University, Shanghai 200438, China\\ \email{song\_zhang@fudan.edu.cn}}
%
%
\maketitle
%

\abstract{\textit{Many probes are proposed to determine the quark-gluon plasma and explore its properties in ultra-relativistic heavy-ion collisions. Some of them are related to initial states of the collisions, such as collective flow, Hanbury-Brown-Twiss (HBT) correlation, chiral magnetic effects and so on. The initial states can come from geometry overlap of the colliding nuclei,  fluctuations or  nuclear structure with the intrinsic geometry asymmetry. The initial geometry asymmetry can transfer to the final momentum distribution in the aspect of hydrodynamics during the evolution of the fireball. Different from traditional methods for nuclear structure study, the ultra-relativistic heavy-ion collisions could  provide a potential platform to investigate nuclear structures with the help of the final-state  observables after the fireball expansion. This chapter first presents a brief introduction of the initial states in relativistic heavy-ion collisions, and then delivers a mini-review for the nuclear structure effects on experimental observables in the relativistic energy domain.}}


\section{\textit{A brief introduction to the relativistic heavy-ion collisions and the initial state}}

Relativistic heavy-ion collisions aim at investigating a new state of matter, quark-gluon plasma (QGP) which was predicted by quantum chromodynamics (QCD)~\cite{QCD-QGP-RHIC} and is considered to be produced at the early stage of central nucleus-nucleus collisions in experiments~\cite{RHICWhitePaper-STAR,FisrtResultsALICE}. The collective motions in partonic level were reported by the RHIC-STAR collaboration~\cite{STAR-NCQv2-PhysRevLett.92.052302}. Recently, a mount of experimental results are obtained to investigate the relationship between the initial states of the collision system and the properties of QGP, such as collective flow~\cite{STARv2BES,PhysRevLett.121.222301-pAlpAuHe3Au}, Hanbury-Brown-Twiss (HBT) correlation~\cite{STAR-HBTBES}, fluctuation~\cite{STAR-FLUCBES} and so on. In theoretical side, the initial geometry fluctuations are studied by various models, such as a multi-phase transport (AMPT) model~\cite{AMPTInitFluc-1}, hydrodynamics~\cite{PhysRevC.100.024904-PbPbXeXeArArOO} etc. Some methods are proposed to perform collective flow and geometry eccentricity analysis related to initial fluctuations~\cite{PartPlane-2}. 

The probes, such as collective flow, HBT correlation or fluctuations,  have been extensively investigated and a new state of matter, so called QGP, was declared to be created in ultra-relativistic heavy-ion collisions, such as in Au + Au or Pb + Pb collisions, as introduced above. Some similar phenomena are also observed in small systems ($p$ + Al, $p$ + Au, $d$ + Au, $^3$He + Au, $p$ + Pb and $p$ + $p$) with high multiplicity events ~\cite{PhysRevLett.121.222301-pAlpAuHe3Au,smallSystemALICE2013} as in large systems (Au + Au and Pb + Pb). It is an open question how to understand transformation coefficient from initial geometry distribution or fluctuation to momentum distribution at final stage in hydrodynamic mechanism~\cite{HydroKN-1,PhysRevC.100.024904-PbPbXeXeArArOO} and if the matter created in different-size  systems undergoes the similar dynamical process and has similar viscosity~\cite{RevModPhys.89.035001-2017}. There are already a lot of theoretical works contributing to physics explanation and analysis method in this subject~\cite{PhysRevC.100.024904-PbPbXeXeArArOO,ZHANG2020135366-PLB2020}. Some detailed introduction and discussion can be found in some recent review articles~\cite{smallSystemReview}.  Suggestions to perform small collision system  experiments could provide great opportunities to uncover  the initial geometry effect in different-size  collision systems.

\begin{figure}[htb]
\center
\includegraphics[angle=0,scale=0.5]{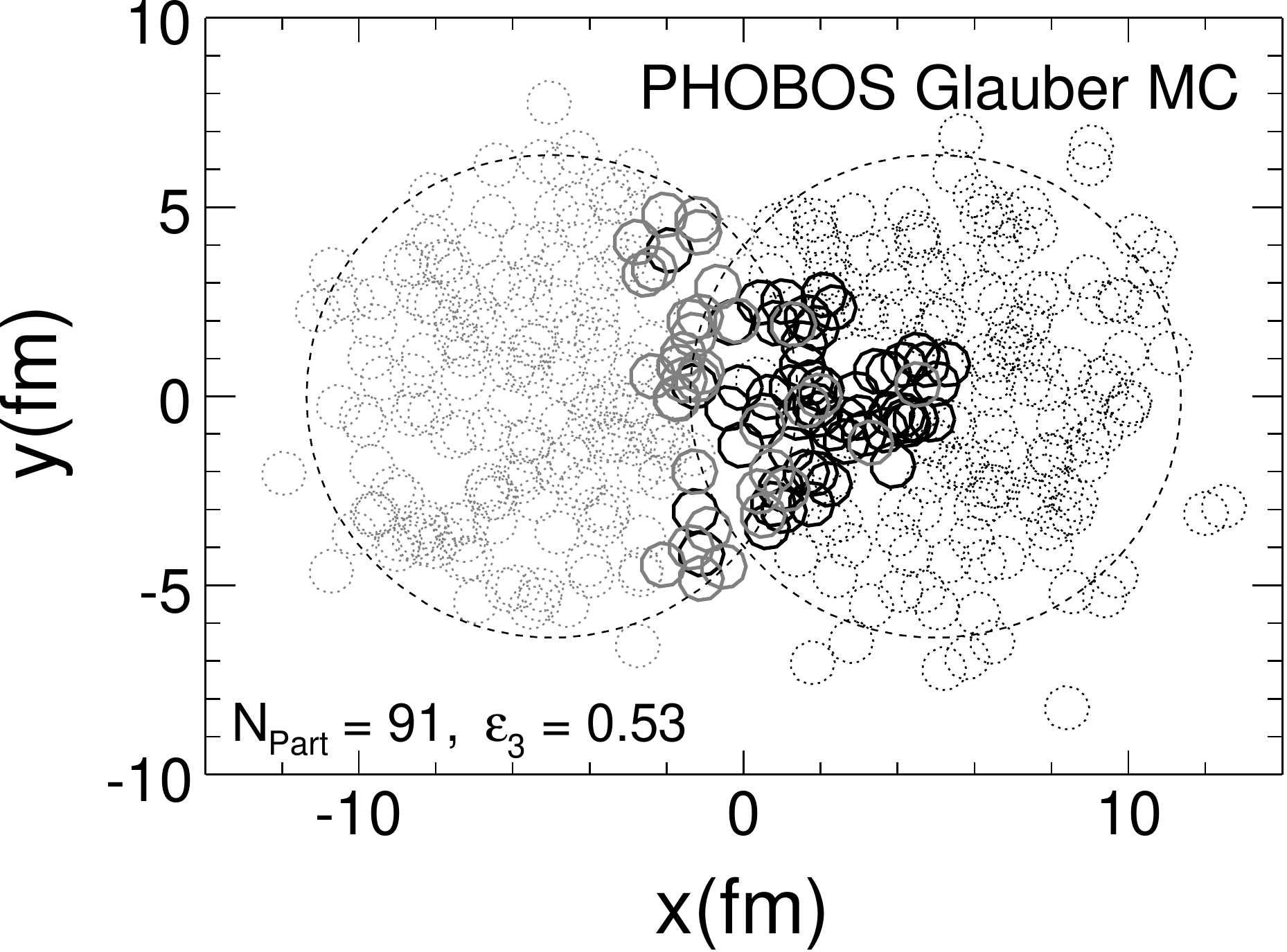}
\caption{Nucleon distribution on the transverse plane in Au + Au collisions at $\sqrt{s_{NN}}$ = 200 GeV simulated by the Glauber Monte Carlo. The nucleons in the two nuclei are shown in gray and black. Wounded nucleons (participants) are indicated as solid circles, while spectators are dotted circles~\cite{PartPlane-2}.
}
\label{fig:PPPlane-Dst}
\end{figure}

The collective flow and the related observables play the key roles in experimental and theoretical investigations, and 
great progress has been made at relativistic energies ~\cite{annurev.nucl.49.1.581-1999}. Systematic measurements of collective flow at center of mass energy from GeV to TeV in nucleus-nucleus collisions have been conducted at RHIC and LHC energies ~\cite{RHICWhitePaper-STAR,FisrtResultsALICE} and the  number of constituent quark (NCQ) scaling law of elliptic flow~\cite{STAR-NCQv2-PhysRevLett.92.052302} becomes one of the significant probes to demonstrate the partonic  level collectivity of the super hot-dense matter. It arises more efforts to understand the transformation of asymmetry from initial geometry space into final momentum space. Specifically the system size dependence of collective flow has been analyzed in experiments~\cite{PhysRevLett.121.222301-pAlpAuHe3Au,smallSystemALICE2013}, and the initial state fluctuations and shear viscosity effect are taken into account in theoretical works~\cite{RevModPhys.89.035001-2017}. The viscous relativistic hydrodynamics models~\cite{RevModPhys.89.035001-2017} suggest a relationship between the collective flow $v_n$ and the initial geometry eccentricity $\varepsilon_n$, $\ln(v_n / \varepsilon_n) \approx -n^2 \langle \frac{\eta}{s}(T)\rangle \langle N_{ch} \rangle ^{-1/3}$, here $\frac{\eta}{s}(T)$ represents the shear viscosity over entropy density, $N_{ch}$ is the number of charge particles. The intuitive picture of the initial state fluctuations is shown in figure~\ref{fig:PPPlane-Dst}~\cite{PartPlane-2}, where the participants contribute to triangularity or higher order geometrical eccentricities, and the triangular flow and higher order flow harmonics are discussed and measured in literature such as  Refs.~\cite{PartPlane-2,PhysRevLett.107.032301-2011-ALICE-vn}. 

Similarly, the number of nucleon (NN) scaling of elliptic flow was first proposed in Ref.~\cite{Yan} by using the quantum molecular dynamics model simulation for lower energy heavy-ion collisions. Even though this conclusion was drawn from low energy heavy-ion collisions but it also works at relativistic heavy-ion collisions \cite{Ko2017}. Actually this NN scaling of elliptic flow  was confirmed by the STAR Collaboration for light nuclei $d$, $t$, $^3He$ (for $\sqrt{s_{NN}}$ = 200, 62.4, 39, 27, 19.6, 11.5, and 7.7 GeV) and anti-deuteron ($\sqrt{s_{NN}}$  = 200, 62.4, 39, 27, and 19.6 GeV) and anti-helium 3 ($\sqrt{s_{NN}}$  = 200 GeV)  \cite{STAR_nnscaling}.

The harmonic coefficients of collective flow  can be calculated from the Fourier series of particle's azimuthal distribution~\cite{PhysRevC.58.1671-FlowMwthod1998},
\begin{eqnarray}
E\frac{d^3N}{d^3p} = \frac{1}{2\pi}\frac{d^2N}{p_Tdp_Tdy}\left(1+\sum_{n=1}^{N}2v_n\cos[n(\phi-\Psi_{RP})]\right),
\label{FlowExpansion}
\end{eqnarray}
where $E$ is the energy, $p_T$ is transverse momentum, $y$ is rapidity, and $\phi$ is  azimuthal angle of the particle. $\Psi_{RP}$ denotes the reaction plane angle. And the Fourier coefficients $v_n (n = 1,2,3,...)$ are collective flows to characterize different orders of azimuthal anisotropies with the form,
\begin{eqnarray}
v_n = \left<\cos(n[\phi-\Psi_{RP}])\right>,
\label{FlowDef}
\end{eqnarray}
where the bracket $\left<\right>$ means statistical averaging over particles and events. The true reaction plane angle $\Psi_{RP}$ is always estimated by event plane angle~\cite{PhysRevC.58.1671-FlowMwthod1998} or by participant plane angle~\cite{PartPlane-2}. The harmonic flow can be calculated with respect to participant plane angle or event plane angle, called as participant plane (PP-) method and event plane (EP-) method, respectively. Some methods avoiding to reconstruct the reaction plane are developed, such as Q-cumulant (QC-) method~\cite{QC-method-1,AMPTInitFluc-1} and two-particle correlation (2PC-) method with rapidity gap~\cite{TwoPartCorrRap-3}. In fact, two-particle correlation method has been already developed earlier in low-intermediate energy heavy-ion collision for flow analysis, eg. see Refs.~\cite{Ma95}.

The collective flow is driven from the initial anisotropy in geometry space. To investigate transformation from initial geometry to final momentum space, the initial geometry eccentricity coefficients $\varepsilon_{n}$ are calculated from the participants via~\cite{AMPTInitFluc-1,PartPlane-2,HydroKN-1},
\begin{eqnarray}
\mathcal{E}_n  \equiv \varepsilon_n \mathrm{e}^{in\Phi_n} \equiv 
  - \frac{\langle r_{part}^{n}\mathrm{e}^{in\phi_{part}}\rangle}
           {\langle r_{part}^{n}\rangle},    
\label{EpsilonPPDef}
\end{eqnarray}
where, $r_{part}$ = $\sqrt{x_{part}^2+y_{part}^2}$ and $\phi_{Part}$ are coordinate position and azimuthal angle of initial participants in the collision zone in the recentered coordinates system ($\langle x_{part}\rangle$ = $\langle y_{part}\rangle$ = 0). $\Phi_n$ is the initial participant plane and $\varepsilon_n$ = $\langle|\mathcal{E}_n|^2\rangle^{1/2}$. The bracket $\langle\rangle$ means the average over the transverse position of all participants event by event. Note that for the definition of  eccentricity coefficients $\varepsilon_{n}$, $r_{part}^{2}$ weight is alternative and it was discussed in Refs.~\cite{HydroKN-1,PartPlane-2}.

From the above introduction, it can be seen that the collective flow and the related observables are sensitive to the initial state in collisions, which may come from the overlap of the collision zone between the collided nuclei, the initial fluctuation as well as the intrinsic geometry of the nuclei. Next the chapter mainly focuses on how the intrinsic geometry (nuclear structure) affects the observables at the final state in the relativistic heavy-ion collisions.

\section{\textit{A brief introduction to the nuclear structure}}

There are a lot of theoretical works (or models) to describe nuclear structure, such as Woods-Saxon model, Shell-model, effective field theory or {\it{Ab~ initio}} calculations. Here  the Woods-Saxon model is only introduced and  a short review is made for the  exotic nuclear structures,  specifically for $\alpha$-clustering nuclei as well as  neutron (proton) skin nuclei in theories and experiments. This is due to that the main purpose of this chapter is to investigate how the exotic nuclear structures affect on final observables in relativistic heavy-ion collisions, or on the other hand how the final observables can be used to distinguish the exotic nuclear structures.


The Woods-Saxon potential was introduced by R. D. Woods and D. S. Saxon~\cite{WoodsSaxon_Dst_1} in 1954 to describe nucleon-nucleus scattering, and written as,
\begin{equation}
    V(r) = \frac{V_0}{1 + e^{(r-R)/a}},
\end{equation}
here $r$ is radial distance of nucleon and $V_0$, $R$, $a$ are different parameters. The ground state of nuclei can be obtained by solving the Schr\"odinger equation by using this potential~\cite{WoodsSaxon_Dst_1_1}.

In relativistic heavy-ion collisions, the nucleon density distribution in the initial collided nuclei can be usually parameterized by a three-parameter Fermi (3pF) distribution~\cite{WoodsSaxon_Dst_2}, namely
\begin{equation}
    \rho(r) = \rho_0\frac{1+\omega(r/R)^2}{1+\exp(\frac{r-R}{a})},
    \label{eq:3pF}
\end{equation}
where $\rho_0$ is the nucleon density in the center of the nucleus, the three parameters $R$, $a$ and $\omega$ are related to the nuclear radius, the (surface) skin depth and deviations from a spherical shape. If $\omega$ sets to zero, it becomes the so-called two-parameter Fermi (2pF) distribution~\cite{EPJC.77.148-2017-NS,PhysRevC.90.014610-NS-BALi,PhysRevLett.112.242502-NSExp1},
\begin{equation}
    \rho(r) = \frac{\rho_0}{1+e^{(r-R)/a}}.
    \label{eq:2pF}
\end{equation}

To describe non-spherical nuclei, the extended formula can be found in  Refs.~\cite{PhysRevC.94.041901-2016-IsobaCME-Deng,PhysRevC.98.054907-2018-Nch-NS} by introducing spherical harmonics and the spatial distribution of deformed nuclei in the rest frame could be written in the Woods-Saxon form,
\begin{equation}
    \rho(r, \theta) = \frac{\rho_0}    {1+\exp\{[r - R_0 - \beta_0 R_0 Y^0_2(\theta) ] / a\}},
    \label{eq:deformedWS}
\end{equation}
where $\rho_0$ is the normal nuclear density, $R_0$ and $a$ represent the ``radius" of the nucleus and the surface diffuseness parameter, respectively, and $\beta_2$ was the deformation of the nucleus.

The Woods-Saxon distribution of nucleons, introduced above, always gives the same radius of proton and neutron distribution in the nucleus, denoted as $R_p$ and $R_n$, respectively. However, if the density profiles are the same for protons and neutrons in neutron-rich nucleus, the radius of neutron distribution should be larger than that of proton's, i.e. $R_n > R_p$, which is called the neutron skin. From the measurements at lower energies, the neutron skin thickness which is defined as the difference between the root mean square radii for the neutron and proton distributions, is presented and  the mechanism is investigated in 
Refs.~\cite{PhysRevLett.112.242502-NSExp1}. To configure a nucleus with neutron skin in a model, one can refer to literature~\cite{PhysRevC.90.014610-NS-BALi}, and here a method based on the two-parameter Fermi (2pF) distribution in equation~(\ref{eq:2pF}) is introduced. For a nucleus with the neutron skin, the nuclear density is written by $\rho^A(r) = \rho^{p,A}(r) + \rho^{n,A}(r)$ and different choice of parameters for protons and neutrons results in the neutron skin effect.

In nature, cluster structure is an ordinary phenomenon that can be observed  everywhere, such as cloud in the sky, water clusters into ice in the river and atom forms crystal. In general, it can be considered the constituents to form a stable structure through interaction each other even though some of the potential among constituents has not been found very well.
Keep in mind that the nuclear structure effects are taken into account in  heavy-ion collisions, or on the other hand, it will be discussed for the investigation of the exotic nuclear structures from the final observables, then a brief introduction to $\alpha$-clustered nuclei will be given for theoretical works and experimental progress. 

\begin{figure}[htb]
\center{
\includegraphics[angle=0,scale=0.25]{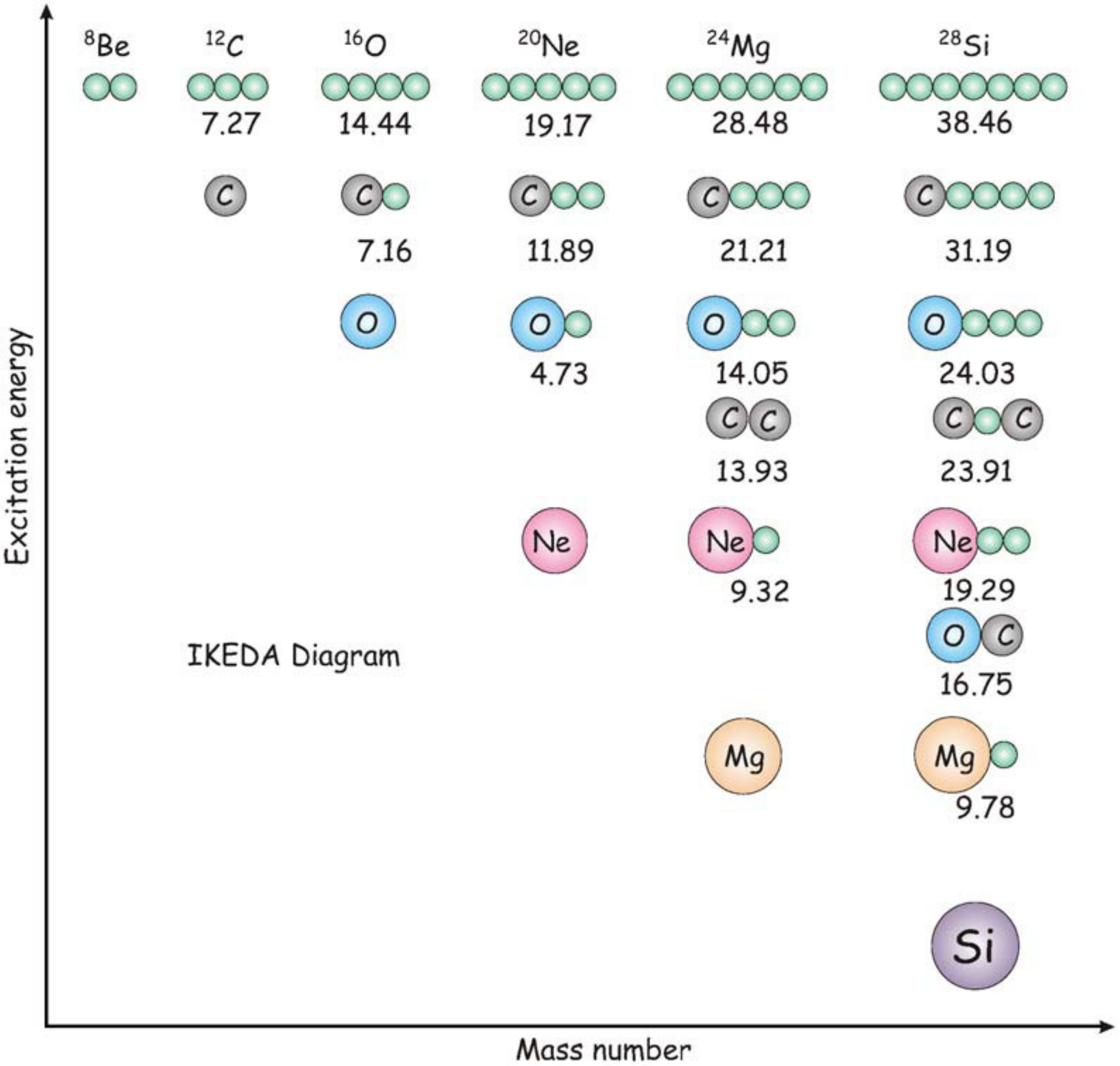}
\caption{Ikeda threshold diagram for $\alpha$-clustered nuclei. The figure is taken  from Ref.~\cite{alpha_plot_VONOERTZEN200643}.}
\label{fig:Ikeda_threshold_diagram}
}
\end{figure}

\begin{figure}[htb]
\center
\includegraphics[angle=0,scale=0.3]{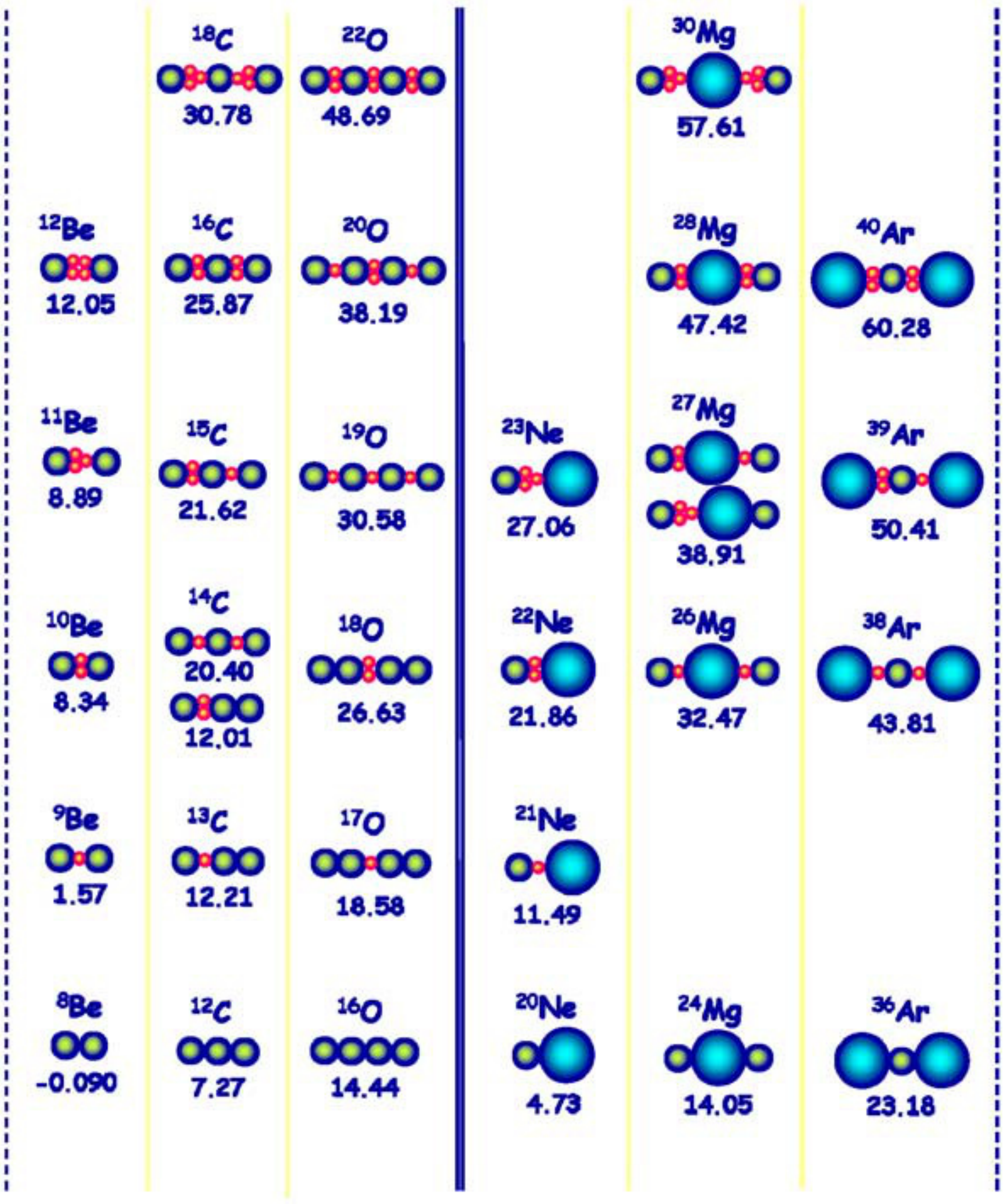}
\caption{Extended threshold diagram. The figure is taken  from Ref.~\cite{alpha_plot_VONOERTZEN200643}.
\label{fig:fig_Extended_threshold_diagram}
}
\end{figure}

The history of clustered nuclei hypothesis which was first proposed by Gamow~\cite{alpha_Gamow_1931} can be traced back decades ago and the light 4$n$ nuclei is considered to be composed of $\alpha$-particles, which is the so-called $\alpha$ cluster model. And then in the last decades, there are amount of works contributing to this exciting subject of exotic nuclear structure~\cite{alpha_plot_VONOERTZEN200643}. Some recent reviews can be found in references~\cite{alpha_plot_VONOERTZEN200643}. Fifty years ago, the famous threshold diagram~\ref{fig:Ikeda_threshold_diagram} was put forwarded, as shown in figure, which pointed out that cluster structures were mainly found close to cluster decay thresholds~\cite{alpha_plot_VONOERTZEN200643}. In this prediction, for example the ground state of $^8$Be would be considered as a bound state of 2$\alpha$s. Later on  this diagram was extended to $N\neq Z$ nuclei  which can be composed of $\alpha$s with valence neutrons in Ref.~\cite{alpha_plot_VONOERTZEN200643}, as shown in figure~\ref{fig:fig_Extended_threshold_diagram}.

From the above introduction, it can be found that there is a chain of clustering nuclei from $\alpha$, $^8$Be, $^{12}$C, $^{16}$O ..., and another chain for $N\alpha$+m$n$ ($N$, $m$ the number of $\alpha$ and neutron) nuclei. Actually the investigation on $\alpha$-conjugate nuclei, such as $^8$Be, $^{12}$C, $^{16}$O, is still  an open question for exotic nuclear structures~\cite{ReviewLight_alphaNucl_Schuck_2017}.

Effort on investigation of $^{12}$C is very impressive and the obtained precise structural data of $^{12}$C provides a benchmark for verifying first-principles calculations. Fermionic molecular dynamics (FMD), anti-symmetrized molecular dynamics (AMD) and covariant density functional theory~\cite{12C_th_Liu_2012} support that ground state of $^{12}$C is in a triangle-like structure with three $\alpha$s. The Brink type THSR-wave function was employed to demonstrate that there is strong two-$\alpha$ correlation in ground state of $^{12}$C~\cite{12C_th_ptep_ptu127}. The experimental evidence was reported for triangular symmetry in $^{12}$C at the ground state~\cite{12C_Exp_PRL_113_012502}. The spectra of giant dipole resonance (GDR) was proposed as fingerprint for $\alpha$-clustering light nuclei by using an Extended Quantum Molecular Dynamics (EQMD)~\cite{EQMD_PRL_113_032506}.

$^{16}$O is another interesting nucleus with an additional $\alpha$ with respect to $^{12}$C. The four $\alpha$s in $^{16}$O are considered to be arranged in regular tetrahedral distribution, which corresponds to the ground state in the Ikeda diagram as in figure~\ref{fig:Ikeda_threshold_diagram}. Chiral nuclear effective field theory~\cite{CNEFT_PRL_112_102501} and covariant density functional theory~\cite{12C_th_Liu_2012} support this regular tetrahedral structure in $\alpha$-clustered $^{16}$O. The algebraic cluster model (ACM)~\cite{ACM_BIJKER2002334,CNEFT_PRL_112_102501} was used to produce the rotation-vibration spectrum of $^{16}$O which is comparable to that in experiment and provides a strong evidence for tetrahedral symmetry in $^{16}$O.  The EQMD calculations presented that $^{16}$O ground state shows the tetrahedral structure and the corresponding characteristic spectra of GDR is comparable to the experimental result ~\cite{EQMD_PRL_113_032506}.

\section{\textit{Influence of $\alpha$-clustering effects}}

It is an open question how the initial geometrical asymmetry transfers to final momentum space and how the intrinsic deformation of nuclei plays an important role for the collective motion properties in relativistic heavy-ion collisions. The $\alpha$-clustered light nuclei are arranged in some special geometry structures with the constituent $\alpha$-clusters, such as $^{12}$C in triangle with 3-$\alpha$s~\cite{12C_th_Liu_2012,12C_th_ptep_ptu127,12C_Exp_PRL_113_012502,EQMD_PRL_113_032506}. 

\begin{figure}[htb]
\center
\includegraphics[angle=0,scale=0.55]{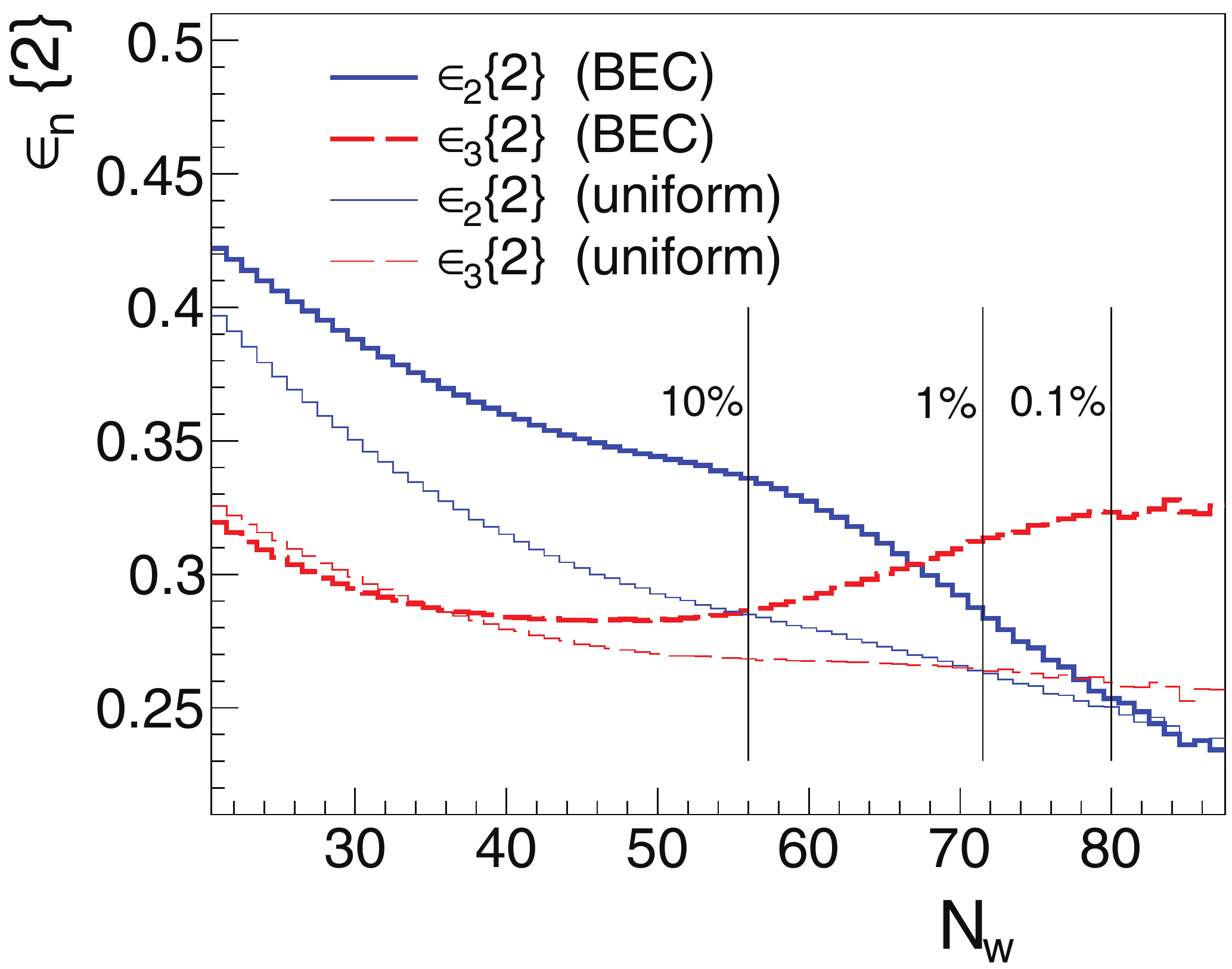}
\caption{The eccentricity simulated by the GLISSANDO model for the Bose-Einstein condensation (BEC) case at RHIC as a function of the wounded nucleons $N_w$ which is corresponding to centralities~\cite{AlphaClusterHIC-Wojciech-PRC}.
}
\label{fig:en-Nw-Broniowski}
\end{figure}

\begin{figure}[htb]
\center
\includegraphics[angle=0,scale=1.5]{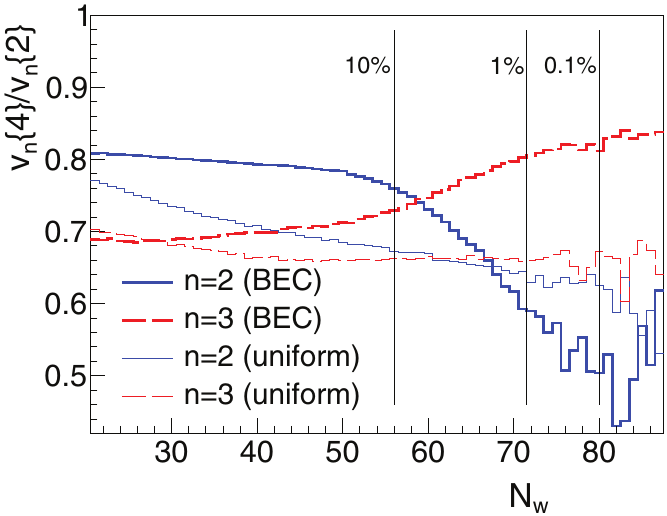}
\caption{Ratios of four-particle to two-particle cumulants plotted as a function of the total number of wounded nucleons for the Bose-Einstein condensation (BEC) case and uniform case~\cite{AlphaClusterHIC-Wojciech-PRC}.
}
\label{fig:vn4-vn2-ratio-Broniowski}
\end{figure}

Refs. ~\cite{AlphaClusterHIC-Wojciech-PRL,AlphaClusterHIC-Wojciech-PRC} proposed an $\alpha$-clustered Carbon colliding against a heavy-ion, $^{12}$C + Au, to investigate the collective flow in the evolution of the fireball. The large deformation in the initial intrinsic nucleus was transformed into the anisotropy of the final momentum space in the fireball. The configuration of the $\alpha$-clustered Carbon is described as following: the centers of three clusters were placed in an equilateral triangle of edge length $l$ and the rms radius of the $\alpha$-cluster is $r_\alpha$. The model parameters $l$ and $r_\alpha$ are optimized by fitting the radial density distribution from the so-called BEC model~\cite{BEC-nucleon-radial-density} and the VMC calculations~\cite{VMC-nucleon-radial-density} as provided at \href{http://www.phy.anl.gov/theory/research/density}{http://www.phy.anl.gov/theory/research/density}. Within the Glauber framework~\cite{GLISSANDO-2}, the so-called mixed model~\cite{Broniowski-init-2} was used to simulate the initial state of the collisions and the event-by-event (3+1)–dimensional viscous hydrodynamics~\cite{Broniowski-hydro} was employed to simulate the evolution of the created fireball. In those works, the initial fluctuations and the intrinsic geometry from $\alpha$-clusters were simulated and discussed. Figure~\ref{fig:en-Nw-Broniowski} from reference~\cite{AlphaClusterHIC-Wojciech-PRC} simulating with clustered and uniform $^{12}$C showed that the geometry enhances the triangularity at high values of the number of wounded nucleons, $N_w$ (corresponding to central collisions), and raises ellipticity at lower values of $N_w$ (corresponding to peripheral collisions). With the help of hydrodynamics~\cite{HydroKN-1}, the ratio of $v_n\{4\}$/$v_n\{2\}$ was proposed as a probe to distinguish the initial $\alpha$-cluster structure of $^{12}$C for the relationship to the initial geometry deformations,
\begin{equation}
\frac{\varepsilon_n\{4\}}{\varepsilon_n\{2\}} \simeq \frac{v_n\{4\}}{v_n\{2\}}, 
\label{FlowEccenRelation}
\end{equation}
where $v_n\{k\}$ or $\varepsilon_n\{k\}$ means \textit{nth}-order harmonic flow coefficients or eccentricity coefficients via \textit{k}-particle cumulant moments as defined in~Ref.~\cite{AlphaClusterHIC-Wojciech-PRC}. Figure~\ref{fig:vn4-vn2-ratio-Broniowski} showed the simulation results of $v_n\{4\}$/$v_n\{2\}$ and  it indicated that the geometric triangularity increased for collisions with a larger number participants, corresponding to central collisions, which was straightforward to measure in ultra-relativistic heavy-ion collisions with standard techniques devoted to analysis of harmonic flow~\cite{AlphaClusterHIC-Wojciech-PRC}.

\begin{figure}[htb]
\center
\includegraphics[angle=0,scale=0.6]{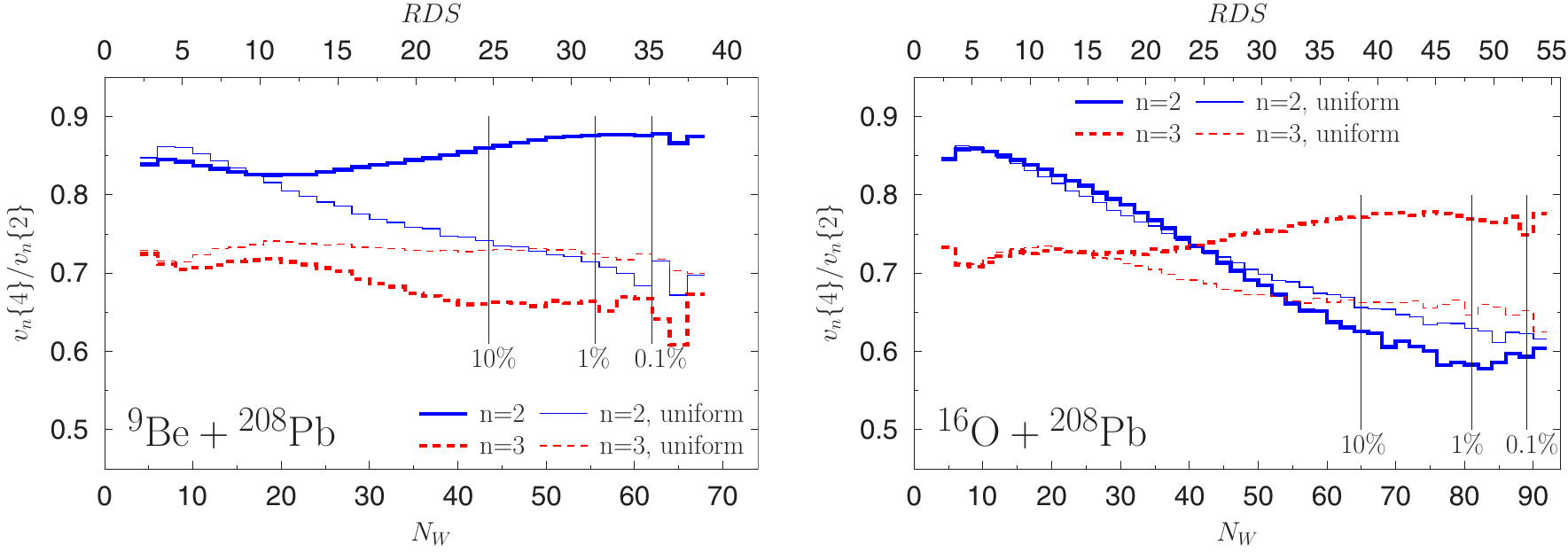}
\caption{Ratios of four-particle to two-particle cumulants for $^{9}$Be + Pb (left) and $^{16}$O + Pb (right) collisions, plotted as a function of the total number of wounded nucleons~\cite{AlphaClusterHIC-Wojciech-PRC-2018}.
}
\label{fig:vn4-vn2-ratio-BePb-OPb-Broniowski}
\end{figure}

Recently the simulations  involving $\alpha$-clustered $^{7,9}$Be, $^{12}$C and $^{16}$O were presented in ultra-relativistic heavy-ion collisions~\cite{AlphaClusterHIC-Wojciech-PRC-2018}. The influence of fluctuation and the intrinsic geometry effect were investigated systematically at the SPS collision energy of $\sqrt{s_{NN}}$ = 17 GeV. In reference~\cite{AlphaClusterHIC-Wojciech-PRC-2018}, the ratios of $v_2\{4\}/v_2\{2\}$ presented increasing (decreasing) trend with the increasing of the wounded nucleons $N_w$ in the collisions with $^{7,9}$Be beams for $\alpha$-clustered (uniform) configuration and the ratios of $v_3\{4\}/v_3\{2\}$ gave the similar trend in $^{16}$O + Pb collisions as shown in figure~\ref{fig:vn4-vn2-ratio-BePb-OPb-Broniowski}. As proposed by Refs.~\cite{LHC-samllsys-propose,RHIC-smallsys-propose}, the collisions with $^{16}$O beam were studied in details~\cite{AlphaClusterHIC-Wojciech-PRC-2018,PhysRevC.102.054907-YALi2020} and were  considered as a platform to invetigate the effect from the intrinsic geometry of nuclei at RHIC or LHC.

The effect from the intrinsic geometry was also studied in a multi-phase transport (AMPT)  model~\cite{AMPT2005,AMPT2021}. AMPT model describes the whole heavy-ion collision processes dynamically as following, the binary collisions among the initial nucleons result in the exited strings and mini-jet partons which all will fragment into partons, and the partons undergo a cascade process till reaching a so-called parton freeze-out status, then the freeze-out partons coalesce into hadrons via a simple coalesence model, and then the hadrons participate in the final rescattering process.
AMPT was extensively used in the community and can successfully describe various physics results  in relativistic heavy-ion collisions at the RHIC and LHC energies~\cite{AMPT2005,AMPT2021}.

During the initialization in the original version of AMPT, the initial nucleon coordinate space in nuclei is described by the HIJING model~\cite{HIJING-2} with the Woods-Saxon distribution~\cite{AMPT2005}. However, the $\alpha$-clustered nuclear structure in the initial condition embodies via reconfiguring in HIJING as following~\cite{PhysRevC.102.054907-YALi2020,EPJA.56.52-JHe2020,PhysRevC.95.064904-SZhang2017}. For three $\alpha$s in equilateral triangle structure placing the $\alpha$ at  each vertice with side length $l_{3\alpha}$ and for four $\alpha$s in tetrahedral structure placing at each vertice with side length $l_{4\alpha}$, for $^{12}$C and $^{16}$O, respectively. The parameters of $l_{3\alpha}$ = 1.8 $fm$ and $l_{4\alpha}$ = 3.42 $fm$ were inherited from the EQMD calculation~\cite{EQMD_PRL_113_032506} and nucleons in the $\alpha$ follow the Woods-Saxon distribution. And the $\alpha$-clustered $^{12}$C and $^{16}$O give the rms radius, 2.47 $fm$ and 2.699 $fm$,  respectively, and the rms radii of $^{12}$C and $^{16}$O configured by the Woods-Saxon distribution were 2.46 $fm$ and 2.726 $fm$, which were all consistent with the experimental data~\cite{ADNDT.99.69-nucleiRMS-ExpData}, 2.47 $fm$ for $^{12}$C and 2.6991 $fm$ for $^{16}$O. Though the $^{12}$C with 3 $\alpha$s in chain structure was unlikely in the ground state, it was also simulated to investigate the initial intrinsic geometry effect, where the 3 $\alpha$s were placed in a line with equivalent distance 2.19 $fm$ giving a rms radius of $^{12}$C 2.47 $fm$.

\begin{figure}[htb]
\center
\includegraphics[angle=0,scale=0.5]{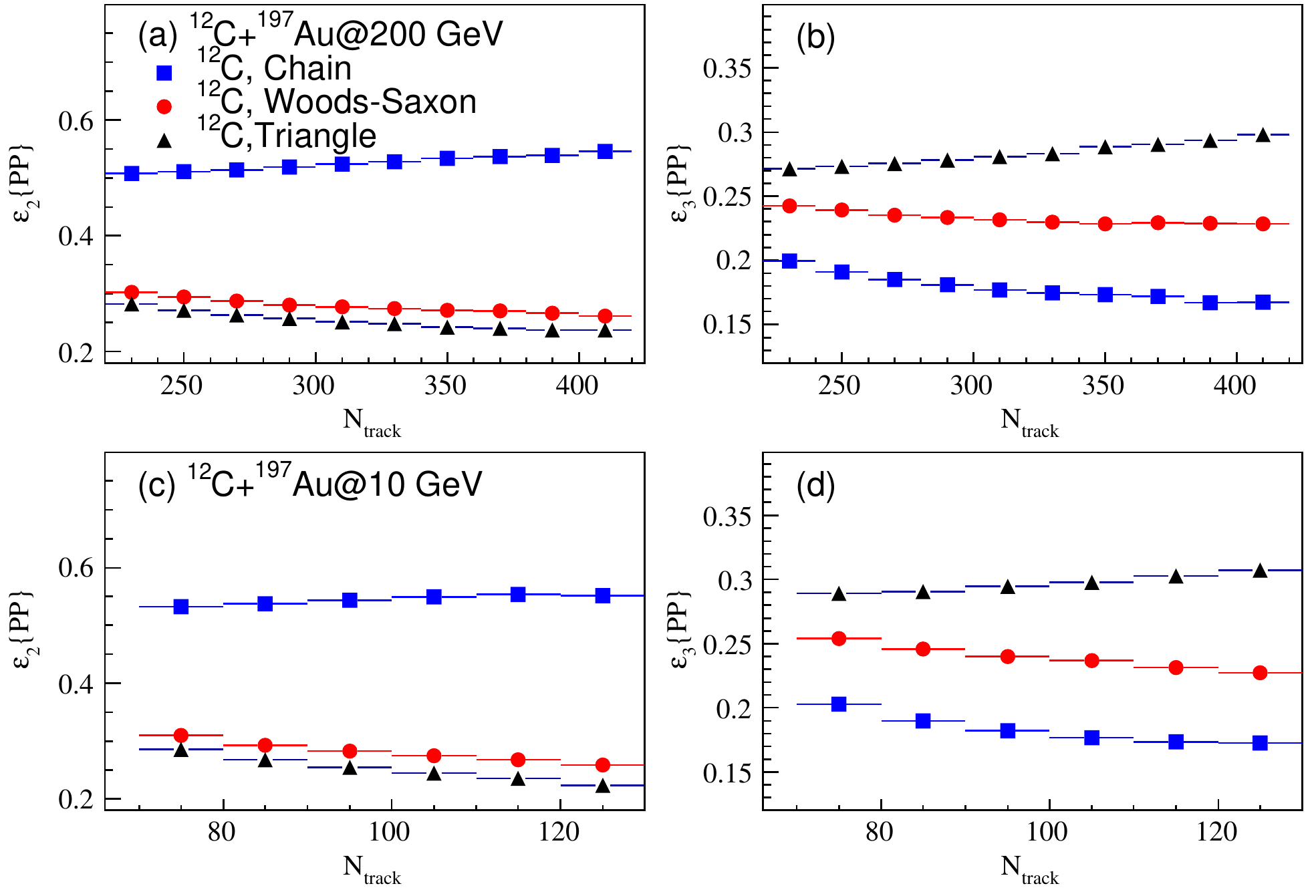}
\caption{In $^{12}\mathrm{C}$ + $^{197}\mathrm{Au}$ collisions at $\sqrt{s_{NN}}$ = 10 GeV and 200 GeV,  second and third order participant eccentricity coefficients, $\epsilon_2\{PP\}$ and $\epsilon_3\{PP\}$, as a function of $\mathrm{N_{track}}$~\cite{PhysRevC.95.064904-SZhang2017}.
}
\label{fig:epsilon-centrality-SZhangYGMa}
\end{figure}

\begin{figure}[htbp]
\center
\includegraphics[angle=0,scale=0.5]{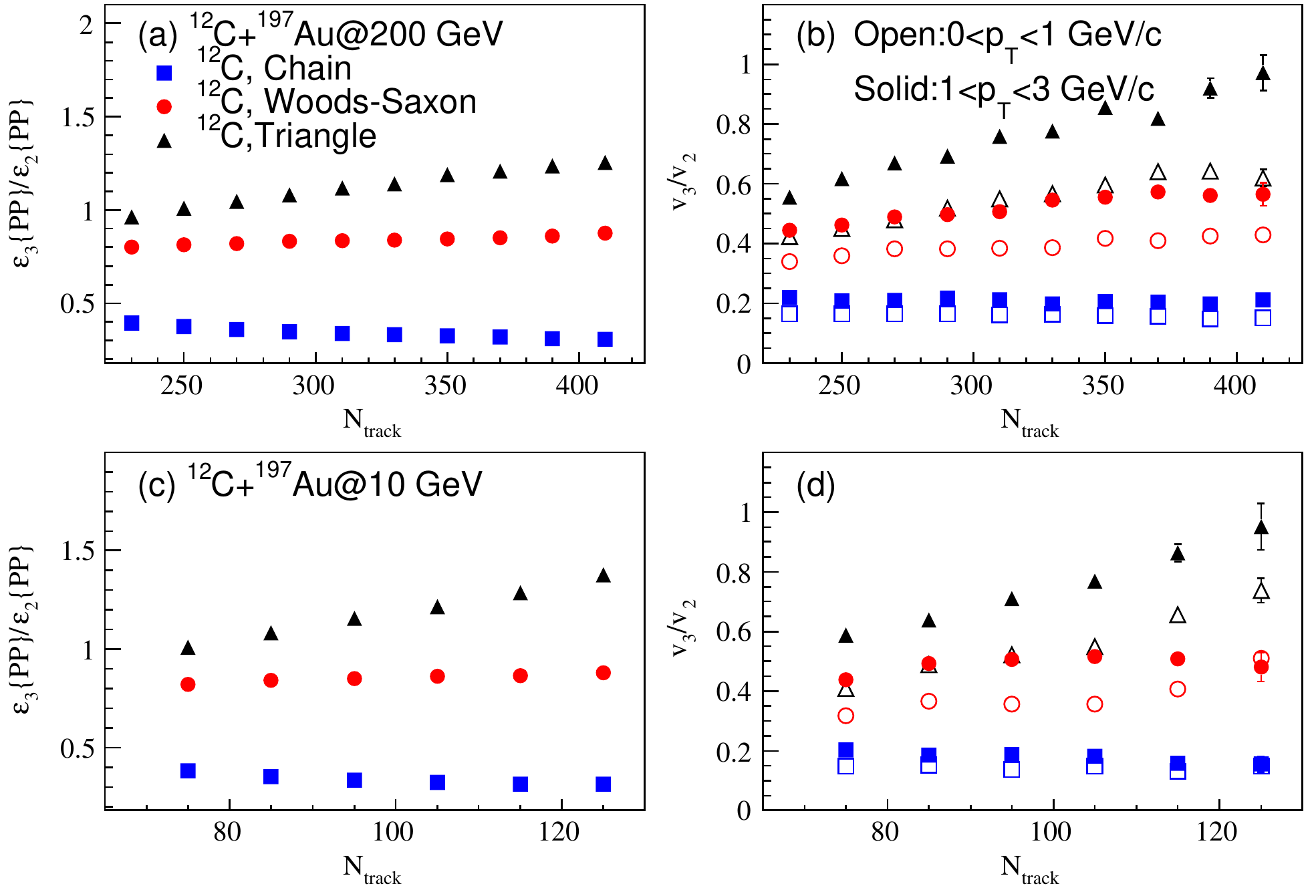}
\caption{
\label{fig:epsilonVn_Ratio-SZhangYGMa}
 In $^{12}\mathrm{C}$ + $^{197}\mathrm{Au}$ collisions at $\sqrt{s_{NN}}$ = 10 GeV and 200 GeV,  $\epsilon_3\{PP\}/\epsilon_2\{PP\}$ and $v_3/v_2$ of final hadrons as a function of $\mathrm{N_{track}}$~\cite{PhysRevC.95.064904-SZhang2017}.
 }
\end{figure}

Once it is possible to provide the reasonable initial stable nuclei with different intrinsic geometry structure, they were introduced to the initial state in HIJING process of AMPT model to investigate the $\alpha$-cluster influence in relativistic heavy-ion collisions. The initial geometry in $^{12}$C + $^{197}$Au collisions was characterised by the eccentricity coefficients $\varepsilon_n$ calculated using equation~(\ref{EpsilonPPDef}). Since the initial fluctuation should be also taken into account, the initial coordinates of participants were used to calculate $\varepsilon_n$ and to reconstruct the so-called participant plane~\cite{AMPTInitFluc-1,PartPlane-2},
\begin{eqnarray}
\Psi_n\{PP\} = \frac{\mathrm{atan2}\left(\left<r^2\sin\left(n\varphi_{part}\right)\right>,\left<r^2\cos\left(n\varphi_{part}\right)\right>\right)+\pi}{n},
\label{PartPlanDef}
\end{eqnarray}
where, $\Psi_n\{PP\}$ is the {\textit n}th-order participant plane angle, $r$ and $\varphi_{part}$ are coordinate position and azimuthal angle of participants in
 the collision zone at initial state, and the average $\left<\cdots\right>$ denotes density weighting.
Then the harmonic flow coefficients with respect to participant plane can  be defined as,
\begin{eqnarray}
v_n \equiv \left<\cos(n[\phi-\Psi_n\{PP\}])\right>.
\label{FlowPPDef}
\end{eqnarray}

Figure~\ref{fig:epsilon-centrality-SZhangYGMa} from Ref.~\cite{PhysRevC.95.064904-SZhang2017} showed the calculated second and third order eccentricity coefficients, $\varepsilon_2\{PP\}$ and $\varepsilon_3\{PP\}$ in $^{12}$C + $^{197}$Au collisions at $\sqrt{s_{NN}}$ = 200 GeV and 10 GeV for different configurations of $^{12}$C, i.e. the $\alpha$-clustered chain and triangle structure as well as nucleon distribution in the Woods-Saxon distribution. At the top RHIC energy ($\sqrt{s_{NN}}$ = 200 GeV), the chain (triangle) structure enhanced the seconded (third) order geometry coefficient and $\varepsilon_2\{PP\}$ ($\varepsilon_3\{PP\}$) increased with the increasing of the number of particles, $N_{track}$, in the collisions. In the Woods-Saxon case, the geometry coefficients ($\varepsilon_2\{PP\}$ and $\varepsilon_3\{PP\}$) slightly decreased with the increasing of $N_{track}$, which indicated the fluctuation contribution gradually weakened from peripheral collisions to central collisions. These initial geometry properties were consistent with those in references~\cite{AlphaClusterHIC-Wojciech-PRL,AlphaClusterHIC-Wojciech-PRC} and  comparable with the presented results in figures~\ref{fig:en-Nw-Broniowski}. With the help of AMPT model, the evolution of collisions was simulated dynamically from the initial nucleon-nucleon  collisions, partonic interaction to the hadronic rescattering and  the initial geometry asymmetry is transferred to the final momentum space. Figure~\ref{fig:epsilonVn_Ratio-SZhangYGMa} presented the ratios of $\varepsilon_3\{PP\}/\varepsilon_2\{PP\}$ and $v_3\{PP\}/v_2\{PP\}$ which  reflected the initial geometry properties: $v_3\{PP\}/v_2\{PP\}$ from triangle structure configuration of $^{12}$C displayed an increasing trend with the increasing of $N_{track}$, that from the  Woods-Saxon configuration kept a flat pattern in a low $p_T$ region and that from the chain structure configuration of $^{12}$C also kept flat pattern and lower than that from the Woods-Saxon configuration. It indicated that the flow coefficient ratio of $v_3\{PP\}/v_2\{PP\}$ was sensitive to the initial geometry properties. Via comparing the results from the  Woods-Saxon case and the $\alpha$-clustered triangle case, the ratio of $v_3\{PP\}/v_2\{PP\}$ could be taken as a probe to distinguish the exotic nuclear structure in heavy-ion collisions, note that the chain structure of $^{12}$ was always considered in excited state.

A light nucleus with the intrinsic geometry structure colliding against a heavy  nucleus can be a feasible experimental project to be performed at RHIC or LHC from the above discussion~\cite{AlphaClusterHIC-Wojciech-PRL,AlphaClusterHIC-Wojciech-PRC,AlphaClusterHIC-Wojciech-PRC-2018,PhysRevC.95.064904-SZhang2017}. The system scan project is also important for investigating the effects from intrinsic geometry and fluctuations, which involves centrality dependence (such as the above introduction), asymmetrical collision systems (light nuclei + different-size nuclei),  and symmetrical collision systems. That work~\cite{PhysRevC.102.054907-YALi2020} proposed three cases of collision systems:  case I, $\mathrm{^{16}O}$ nucleus (with or without $\alpha$-cluster) +  ordinary nuclei (always in the Woods-Saxon distribution) in the most central collisions;  case II,  $\mathrm{^{16}O}$ nucleus (with or without $\alpha$-cluster)  + $\mathrm{^{197}Au}$ nucleus collisions for centrality dependence, and  case III, symmetric collision systems (namely, $^{10}$B + $^{10}$B, $^{12}$C + $^{12}$C, $^{16}$O +  $^{16}$O (with or without $\alpha$-cluster),  $^{20}$Ne +  $^{20}$Ne,  and  $^{40}$Ca + $^{40}$Ca) in the most central collisions. In that work the linear mode in the initial geometry eccentricity coefficients and flow coefficients were calculated~\cite{PhysRevC.102.054907-YALi2020,ZHANG2020135366-PLB2020}. And for higher-order initial geometry eccentricity coefficients and flow coefficients, the relationship for linear mode can be described by $v_n^L$ $\propto$ $\varepsilon_n^L$ (see Refs.~\cite{ZHANG2020135366-PLB2020} and references therein for details).

\begin{figure*}[htb]
\center
	\includegraphics[angle=0,scale=0.6]{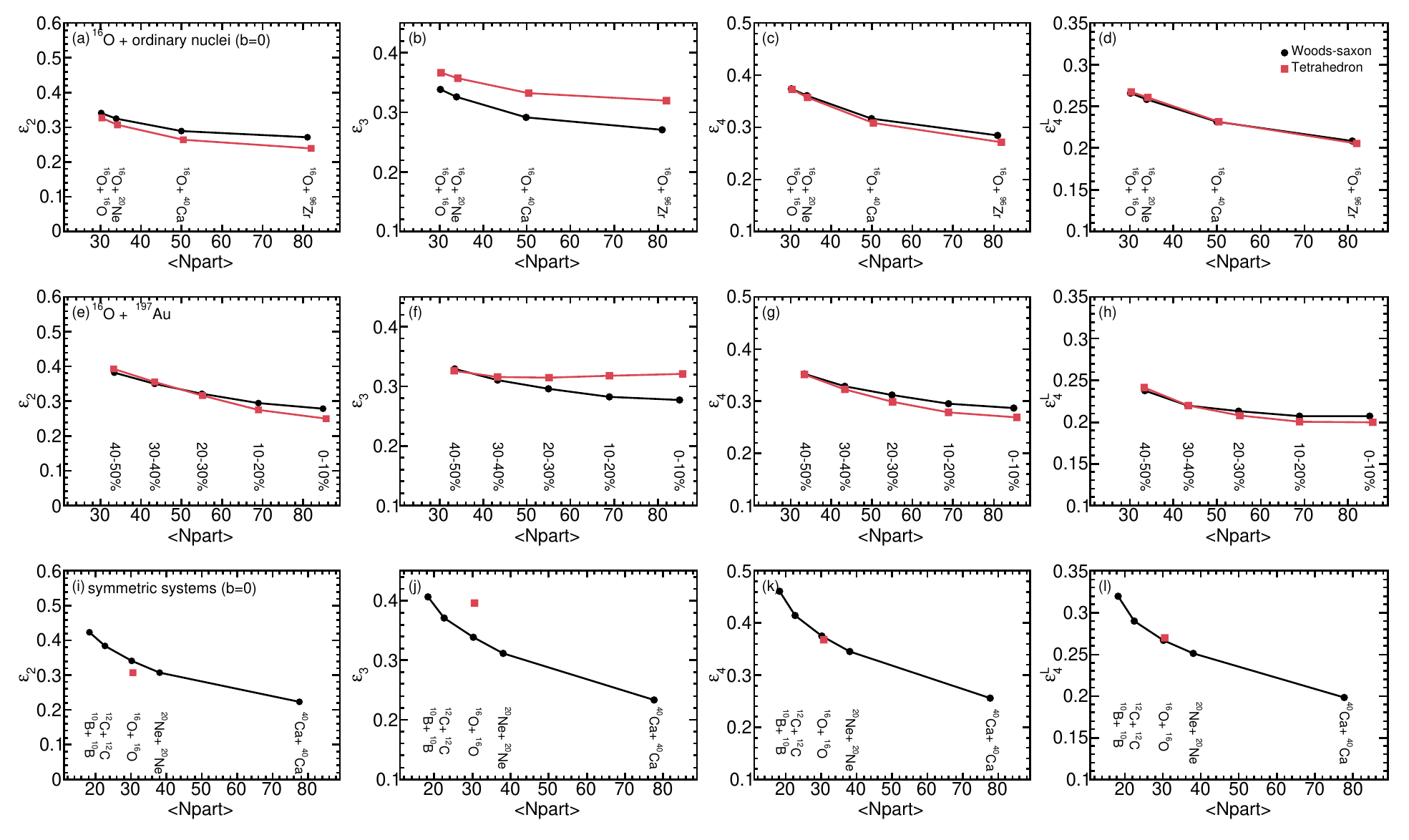}
	\caption{Eccentricity coefficients, namely $\varepsilon_2$, $\varepsilon_3$, $\varepsilon_4$ and $\varepsilon_4^L$ (from left column to right column) as a function of number of participants $\left<N_{part}\right>$. 
	Upper panels (Case I: (a), (b), (c), and (d)) are the results from  the most central collisions of the $\mathrm{^{16}O}$ nucleus (with or without $\alpha$-cluster) + ordinary nuclei ($^{16}$O, $^{20}$Ne, $^{40}$Ca and $^{96}$Zr), middle panels (Case II: (e), (f), (g), and (h)) are the results for the centrality dependence of the $\mathrm{^{16}O}$ (with or without $\alpha$-cluster) + $^{197}$Au collisions, and lower panels (Case III: (i), (j), (k), and (l)) represent the symmetric collision systems from small  to large ones in the most central collisions, i.e.  $^{10}$B + $^{10}$B, $^{12}$C + $^{12}$C, $^{16}$O +  $^{16}$O (with or without $\alpha$-cluster),  $^{20}$Ne +  $^{20}$Ne,  and  $^{40}$Ca + $^{40}$Ca.~\cite{PhysRevC.102.054907-YALi2020}}
	\label{fig:enL-YALi}
\end{figure*}

\begin{figure*}[htb]
\center
	\includegraphics[angle=0,scale=0.6]{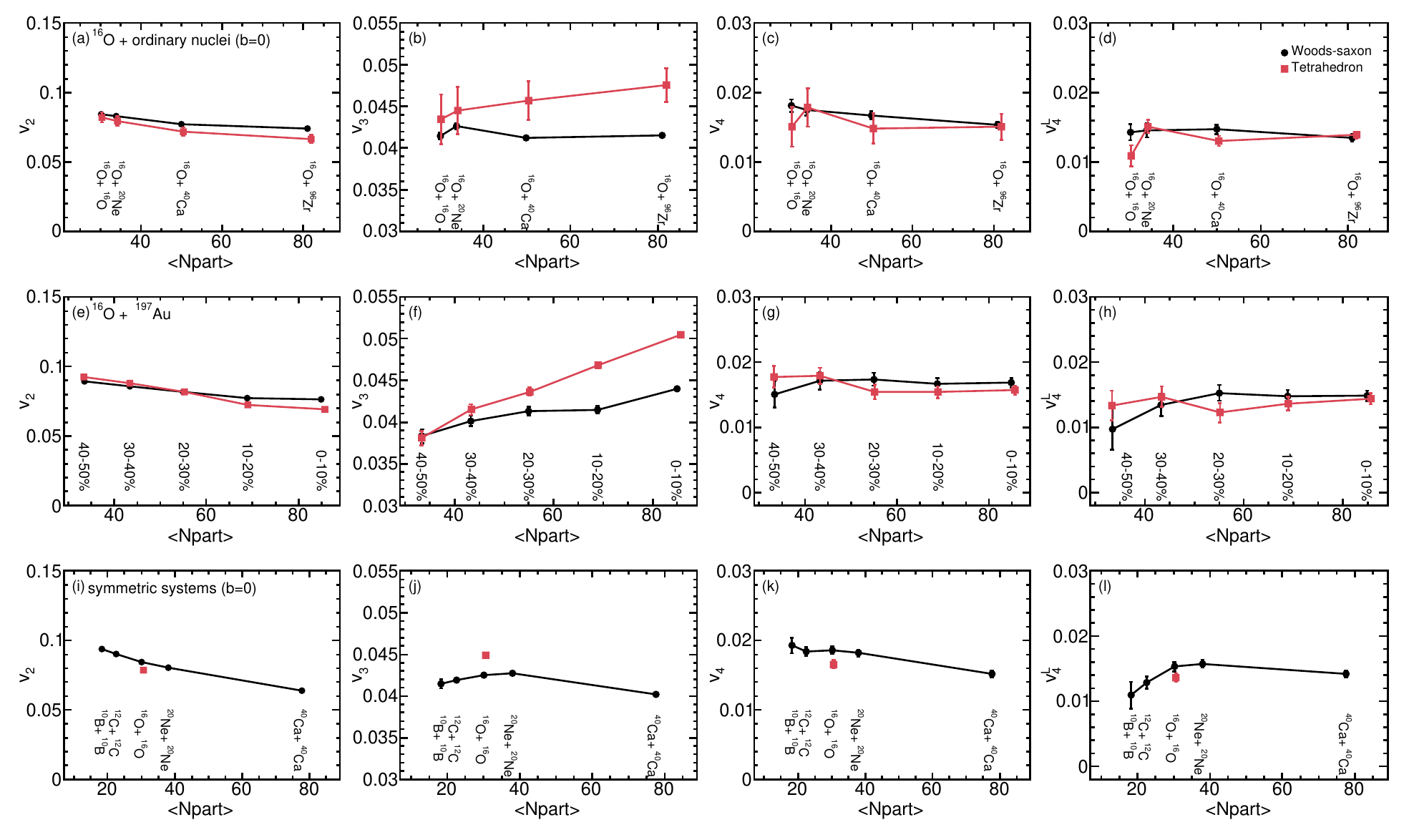}
	\caption{Same as Fig.~\ref{fig:enL-YALi} but for anisotropic flow coefficients, namely $v_2$, $v_3$, $v_4$ and its linear mode  $v_4^L$.~\cite{PhysRevC.102.054907-YALi2020}}
	\label{fig:vnL-YALi}
\end{figure*}

\begin{figure*}[htb]
\center
	\includegraphics[angle=0,scale=0.6]{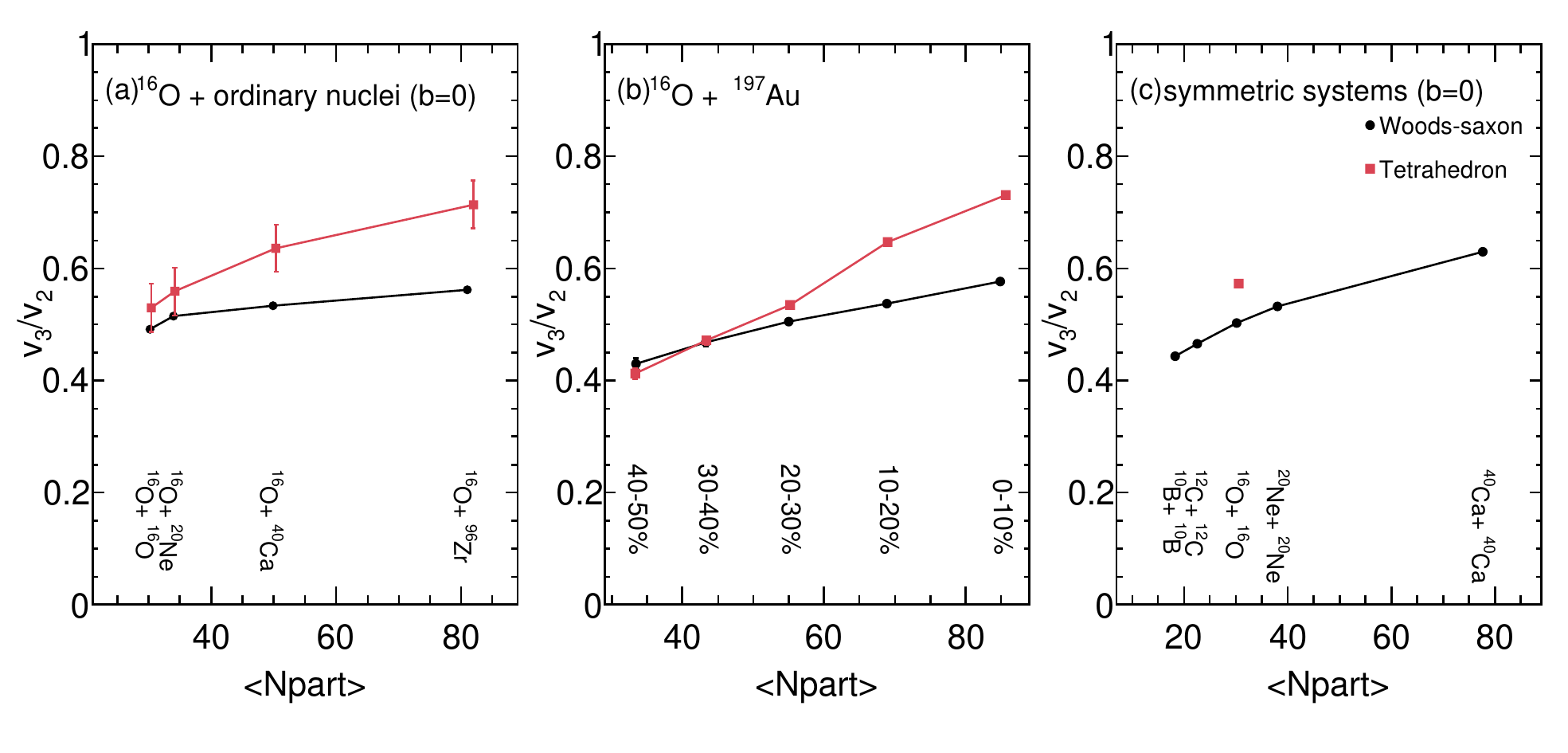}
	\caption{Ratio of $v_3/v_2$ as a function of $\left<N_{part}\right>$:  the case I, i.e.  $\mathrm{^{16}O}$  + ordinary nuclei at b = 0 fm (a), the case II,  i.e. $\mathrm{^{16}O}$ + Au at different centralities (b), the case III, i.e. symmetric collisions at b = 0 fm (c). The red or black lines (symbols) represent $^{16}$O w/ or w/o $\alpha$-cluster structure.~\cite{PhysRevC.102.054907-YALi2020}}
	\label{fig:v3v2-ratio-YALi}
\end{figure*}

By using the AMPT model, the three-case collision systems were simulated and the initial geometry coefficients and the final flow coefficients are presented in figure~\ref{fig:enL-YALi} and~\ref{fig:vnL-YALi},  respectively~\cite{PhysRevC.102.054907-YALi2020}. From figure~\ref{fig:enL-YALi}, $\varepsilon_2$, $\varepsilon_3$ and $\varepsilon_4^L$ kept the decreasing trend with the increasing of the $\langle N_{part}\rangle$ (number of participants corresponding to centrality) which confirmed that the fluctuation contribution became less in  central collisions or in larger collisions system. $\varepsilon_3\{PP\}$ in the collision systems with tetrahedron structure $^{16}$O beam was different from that in the Woods-Saxon case, as shown in panel (b), (f) and (j) in figure~\ref{fig:enL-YALi}, and $\varepsilon_2\{PP\}$, $\varepsilon_4\{PP\}$ and $\varepsilon_4^L\{PP\}$ presented minor difference for the different configurations of $^{16}$O beam. It was clear that $\varepsilon_3\{PP\}$ in the collision system with tetrahedron $\alpha$-clustered $^{16}$O beam deviated from the system dependence from $^{10}B$ + $^{10}B$, $^{12}$C + $^{12}$C, $^{20}$Ne + $^{20}$Ne to $^{40}$Ca + $^{40}$Ca collisions on panel (j) in figure~\ref{fig:vnL-YALi}. After undergoing the dynamical evolution in AMPT via partonic interaction and hadronic rescattering, the initial geometry properties were still survived and transferred to the momentum space by analyzing the flow coefficients as shown in figure~\ref{fig:vnL-YALi}. If comparing $\varepsilon_n\{PP\}$ ($\varepsilon_4^L\{PP\}$) with $v_n$ ($v_4^L$), one can find that the flow coefficients which characterised the momentum asymmetry reflected the initial geometry coefficients which denoted the initial state of the collision system. To further compare three cases, the ratio of anisotropic flow coefficients $v_3/v_2$ was presented in figure~\ref{fig:v3v2-ratio-YALi} which all presented difference from the configuration of $^{16}$O beam, namely $\alpha$-clustered type gave the nontrivial system dependence in comparing with the Woods-Saxon type. 
It indicated that the case I and case III for  the most central collisions could reduce the geometry distribution influence from the overlap region as much as possible, and then it suggested that $v_3/v_2$ could be taken as a probe to identify the $\alpha$-clustered structure of $^{16}$O. Therefore the cases I and III were proposed as a potential scenario of a system scan experiment project at RHIC or LHC.

\begin{figure*}[htb]
\center
	\includegraphics[angle=0,scale=0.42]{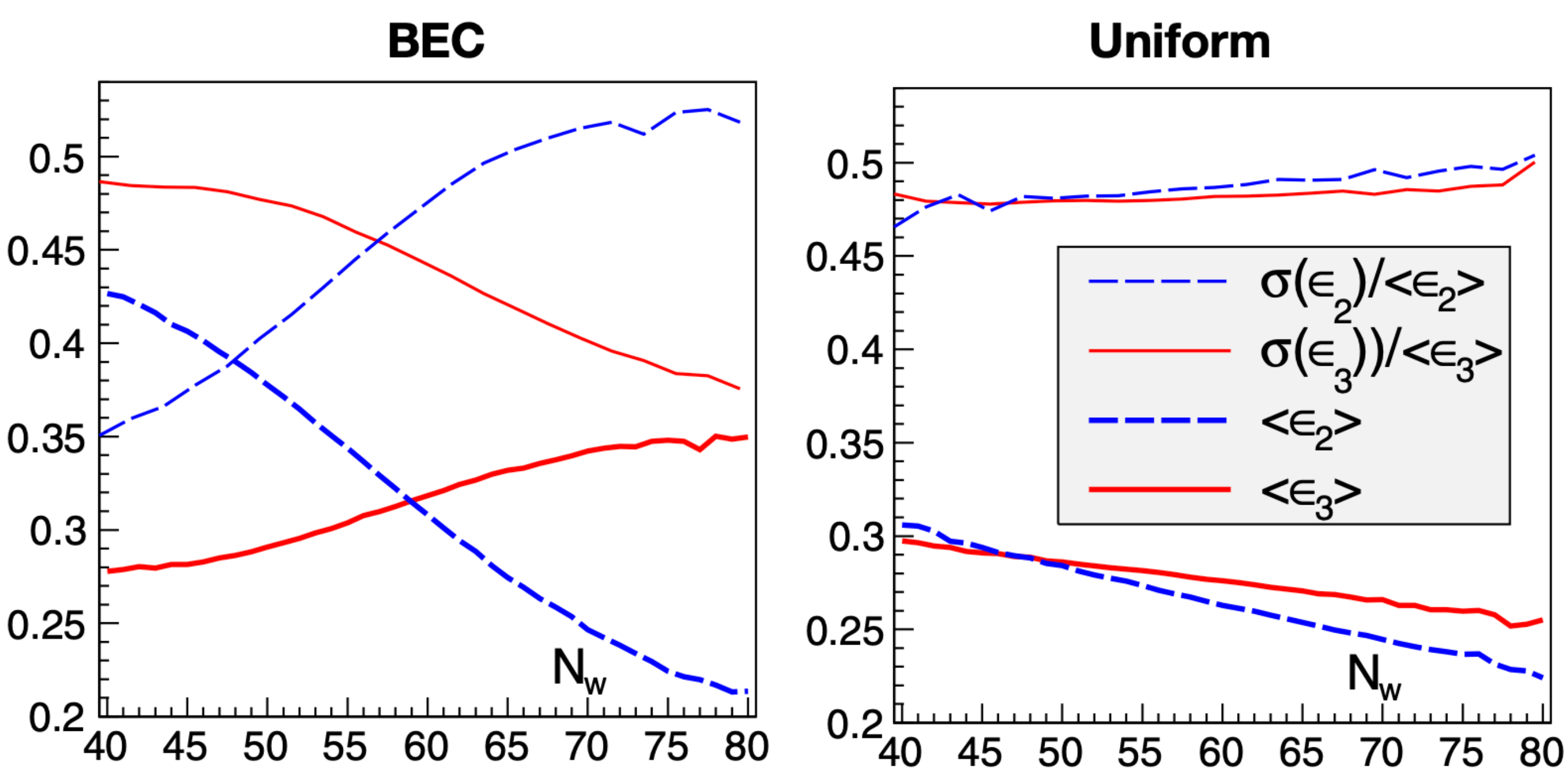}
	\caption{The event-by-event statistical properties of the fireball (average ellipticity, triangularity, and their scaled standard deviations) as a function of the number of wounded nucleons. Left for the BEC case and  right for the  uniform case~\cite{AlphaClusterHIC-Wojciech-PRL}.}
	\label{fig:en_fluctuation-Broniowski}
\end{figure*}

After 
the influence from the intrinsic geometry of the $\alpha$-clustered light nuclei beam was introduced in relativistic heavy-ion collisions via analyzing the transformation from the initial eccentricity to the final collective flow in the above paragraph, 
the fluctuation method is discussed for determining how it is sensitive to the intrinsic geometry and the system size (larger fluctuation in small systems) or to both. The authors~\cite{AlphaClusterHIC-Wojciech-PRL} presented the event-by-event statistical properties of the fireball via the observables of average ellipticity $\varepsilon_2$, triangularity $\varepsilon_3$, and their scaled standard deviations $\sigma(\varepsilon_2)/\varepsilon_2$, $\sigma(\varepsilon_3)/\varepsilon_3$ as a function of the number of wounded nucleons $N_w$ as shown in figure~\ref{fig:en_fluctuation-Broniowski}. In uniform (unclustered) case the scaled standard deviations $\sigma(\varepsilon_2)/\varepsilon_2$, $\sigma(\varepsilon_3)/\varepsilon_3$ presented ignorable $N_w$ dependence with similar value of specific centrality as on right panel in figure~\ref{fig:en_fluctuation-Broniowski}. In BEC (clustered) case, however, $\sigma(\varepsilon_2)/\varepsilon_2$ and  $\sigma(\varepsilon_3)/\varepsilon_3$ taken on the opposite $N_w$ dependence trend. The flow fluctuation quantified with scaled standard deviation should be approximately equal to the eccentricity fluctuation quantified in the same way,
\begin{eqnarray}
\frac{\sigma(v_n)}{\langle v_n\rangle} \approx \frac{\sigma(\varepsilon_n)}{\langle \varepsilon_n\rangle}.
\label{vnen_fluctuation_relation}
\end{eqnarray}
If the relationship was taken into account, it can be concluded that the eccentricity coefficients and the scaled standard deviations were sensitive to the initial geometry properties and could be measured with the standard flow analysis methods.

\begin{figure*}[htb]
\center
	\includegraphics[angle=0,scale=0.42]{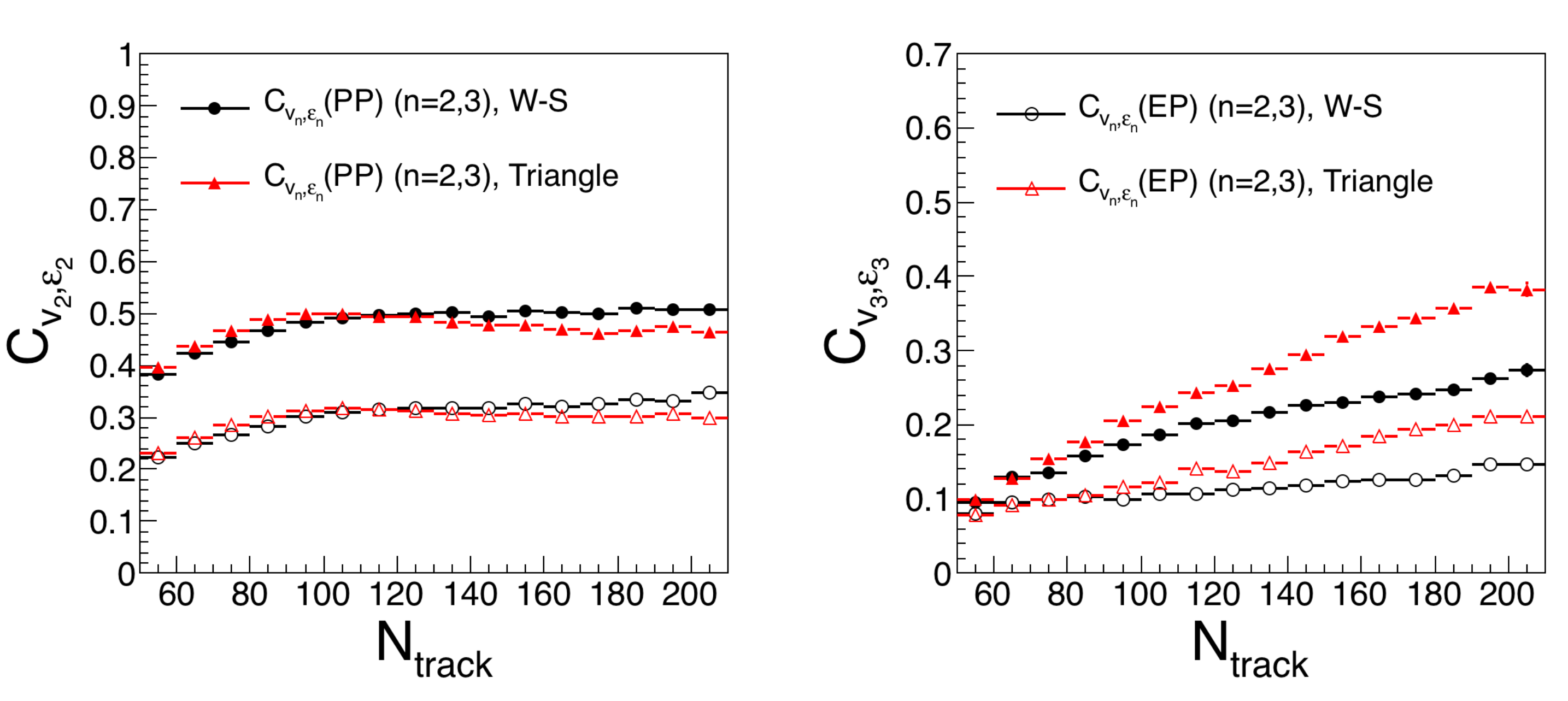}
	\caption{Correlation coefficients $C_{v_n, \varepsilon_n}$ $(n=2,3)$ as a function of $N_{track}$ in $^{12}$C+$^{197}$Au collisions with different geometry configurations of $^{12}$C~\cite{PhysRevC.102.014910-envnFluc-LMa}.}
	\label{fig:envn_correlation_LMa}
\end{figure*}

In reference~\cite{PhysRevC.102.014910-envnFluc-LMa}, fluctuation and correlation of the azimuthal anisotropies in $^{12}$C + $^{197}$Au collisions at $\sqrt{s_{NN}}$ = 200 GeV were explored by using AMPT model. Properties of the initial eccentricity and final harmonic flow fluctuation in the collisions with $^{12}$C beam via the Woods-Saxon configuration and triangular $\alpha$-clustering configuration were investigated via the scaled variance (scaled standard deviations), skewness, and kurtosis,
\begin{eqnarray}
\begin{aligned}
R_{\varepsilon_n} & = \frac{\sigma(\varepsilon_n)}{\langle \varepsilon_n \rangle}=\sqrt{\frac{\langle \varepsilon_n^2 \rangle - \langle \varepsilon_n \rangle^2}{\langle \varepsilon_n \rangle^2}},\\
S_{\varepsilon_n} & = \frac{\langle (\varepsilon_n - \langle\varepsilon_n\rangle)^3 \rangle} {\langle (\varepsilon_n - \langle\varepsilon_n\rangle)^2 \rangle^{3/2}},\\
K_{\varepsilon_n} & = \frac{\langle (\varepsilon_n - \langle\varepsilon_n\rangle)^4 \rangle} {\langle (\varepsilon_n - \langle\varepsilon_n\rangle)^2 \rangle^{2}} - 3.
\end{aligned}
\label{vnen_variance_skewness_kurtosis}
\end{eqnarray}
The flow fluctuations were defined in the same way as equation (\ref{vnen_variance_skewness_kurtosis}). In that work, it was found that $R_{\varepsilon_3}$ and $R_{v_3}$ were sensitive to the triangle structure of $^{12}$C which was consistent with the conclusion in reference~\cite{AlphaClusterHIC-Wojciech-PRL} since there was different in the exact value from two different models. The skewness and kurtosis of the eccentricity fluctuation, namely $S_{\varepsilon_n}$ and $K_{\varepsilon_n}$ presented a significant difference in both magnitude and trend as a function of $N_{track}$ for third-order, however, this difference disappeared for flow fluctuation, namely $S_{v_n}$ and $K_{v_n}$. It was argued that could be from the source evolution that in the final state of AMPT the $v_n$ ($n = 2,3$) distributions were nearly Gaussian thus the skewness and kurtosis of elliptic and triangular flow fluctuations especially at large $N_{track}$ are almost consistent with zero within statistical uncertainties~\cite{PhysRevC.102.014910-envnFluc-LMa}. The correlations of the initial eccentricities with final flow harmonics were investigated via the Pearson coefficient to quantify the strength of the correlation defined as,
\begin{eqnarray}
C_{v_n,\varepsilon_n} = \frac{\langle v_n \varepsilon_n \cos(n[\Psi - \Psi\{PP\}])} {\sqrt{\langle |\varepsilon_n|^2 \rangle \langle |v_n|^2 \rangle}},
\label{vnen_correlation}
\end{eqnarray}
where $\Psi\{PP\}$ was the participant plane angle, $\Psi$ was the phase of the flow coefficient $v_n$ (e.g., $\Psi$ was the event-plane angle if $v_n$ was calculated based on the event-plane method). Figure~\ref{fig:envn_correlation_LMa} showed the calculated $C_{v_n,\varepsilon_n}$ as a function of $N_{track}$ in $^{12}$C + $^{197}$Au collisions with different geometry configuration of $^{12}$C~\cite{PhysRevC.102.014910-envnFluc-LMa}. The increasing trend of $C_{v_n,\varepsilon_n}$ with the increasing of $N_{track}$ indicated stronger linear $v_n-\varepsilon_n$ correlation at larger $N_{track}$ (central collisions). In comparison, the correlation between $v_2$ and $\varepsilon_2$ presented comparable results in the Woods-Saxon and triangle $\alpha$-clustered configuration whereas triangle $\alpha$-clustered configuration gave stronger $v_3-\varepsilon_3$ correlation than the Woods-Saxon configuration especially at large $N_{track}$.

The photon flow observable from the $\alpha$-clustered $^{12}$C and $^{197}$Au collisions could be another potential probe to investigate the initial geometry properties~\cite{EPJA.57.134-PDasgupta2021}. By using a hydrodynamic model  with different structure configurations of $^{12}$C beam, the transverse momentum spectra and anisotropic flow coefficients $v_n$ of thermal photons were calculated in $^{12}$C + $^{197}$Au collisions at $\sqrt{s_{NN}}$ = 200 GeV. Comparing the results from unclustered with $\alpha$-clustered $^{12}$C, the work~\cite{EPJA.57.134-PDasgupta2021} found that the slope of thermal photon spectra was not sensitive to the orientations of collisions, however, the elliptic ($v_2$) and triangular flow ($v_3$) coefficients of direct photons for $\alpha$-clustered $^{12}$C structure were significantly larger and predominantly formed by the QGP radiation. Also a strong anti-correlation between initial spatial ellipticity and triangularity was observed.

In another study at low energy \cite{ShiCZ1}, the $\gamma$-photon energy spectrum is reconstructed, which is in good agreement with the experimental data of $^{86}$Kr + $^{12}$C at $E/A$ = 44 MeV within the framework of the modified EQMD model. The directed and elliptic flows of free protons and direct photons were investigated by considering the $\alpha$-clustering structure of $^{12}$C. The difference in the collective flows between different configurations of $^{12}$C is observed. As the previous work~\cite{EPJA.57.134-PDasgupta2021}, this study also indicated that the collective flows of direct photons are sensitive to the initial configuration. 

As a short summary for the above parts, the intrinsic geometry from $\alpha$-cluster arrangement in light nuclei $^{7,9}$Be, $^{12}$C and $^{16}$O aroused people's attention to investigate the effect from the initial geometry asymmetry and fluctuations and further to understand how the initial geometry properties transfer to the final momentum space. It was found the $\alpha$-clustered geometry structure contributed to not only the collective flows, especially to triangular flow $v_3$, but also to the flow fluctuations, which was sensitive to the initial geometry characteristic.

Additionally, the effect of $\alpha$ cluster  in relativistic collisions via di-hadron azimuthal correlation was also present \cite{WangYZ}. A collision system scan involving $\alpha$-clustered $^{12}$C and $^{16}$O is studied by using a multiphase transport model in the most central collisions at $\sqrt{s_{NN}} = 6.37$ TeV. By comparing the correlation functions of different configurations and extracting root-mean-square width and kurtosis in the away-side region, 
	the results show substantial distinction in correlation functions between Woods-Saxon distribution and $\alpha$-clustered structures. In indicates  di-hadron azimuthal correlation could be seen a potential probe to distinguish $\alpha$-clustered nuclei.

\begin{figure*}[htb]
\center
	\includegraphics[angle=0,scale=0.5]{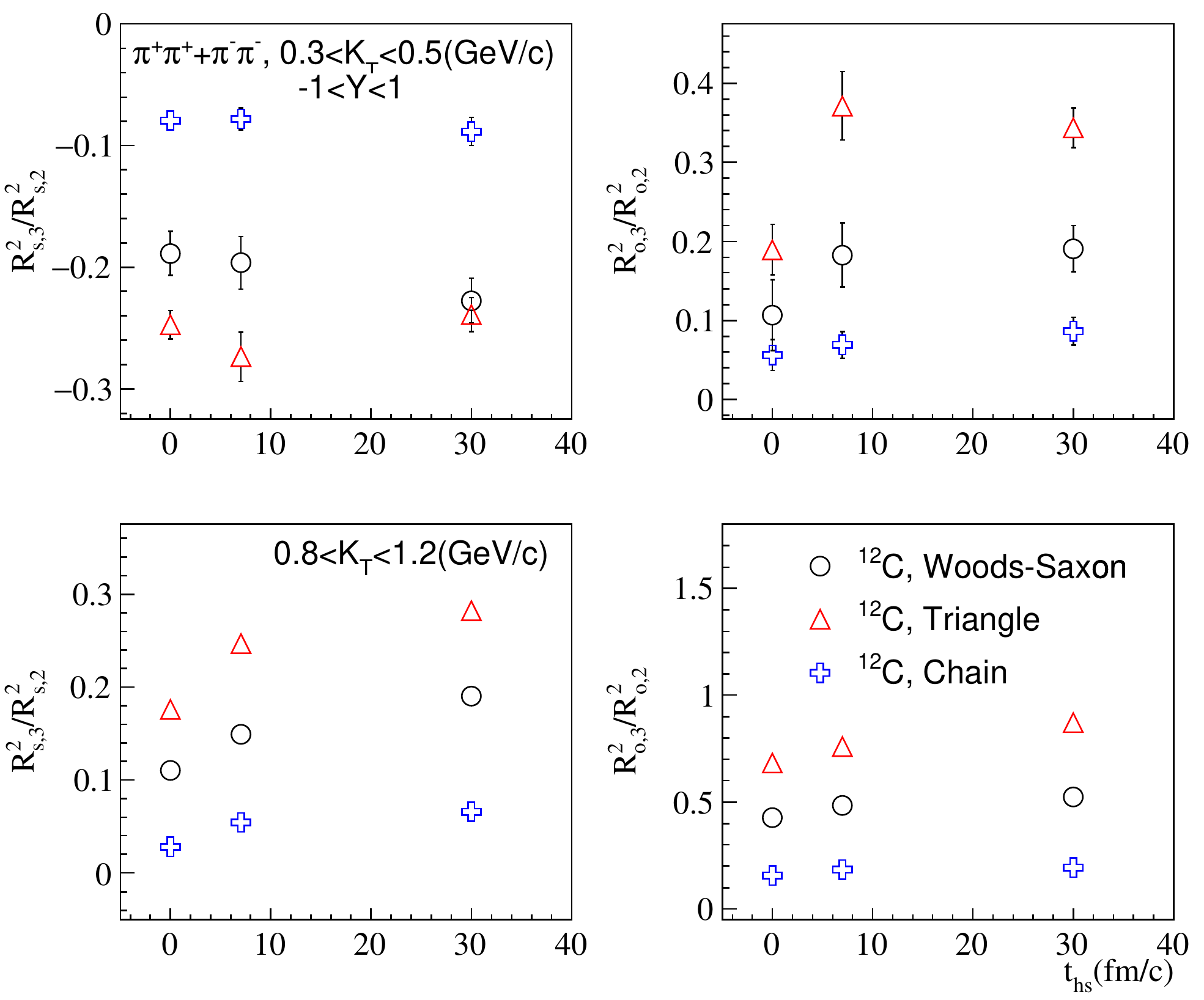}
	\caption{The time evolution of the ratio of the third- to the second-order HBT radii $R^2_{s(o),3}/R^2_{s(o),2}$ with two $K_T$ bins~\cite{EPJA.56.52-JHe2020}.}
	\label{fig:radiiRatio-JHe}
\end{figure*}

Two-particle intensity interferometry was proposed and developed by Hanbury Brown and Twiss (HBT) in the 1950s~\cite{brown1956test} and throughout the last decades, the HBT method had been extensively applied in heavy-ion collisions to study the evolution of the fireball~\cite{annurev.nucl.55-2005-Lisa}. Via the HBT method, two-particle momentum correlation provided a unique way to obtain direct information about the space-time structure. By using AMPT model, the azimuthal angle dependences of the HBT radii relative to the second- and third-order participant plane from $\pi-\pi$ correlations were calculated in $^{12}$C + $^{197}$Au collisions at $\sqrt{s_{NN}}$ = 200 GeV. And the Woods-Saxon distribution, chain and triangle $\alpha$-clustering configuration were taken into account and the comparison among the three cases was discussed in reference~\cite{EPJA.56.52-JHe2020}. It was proposed the ratio of the third- to the second-order HBT radii $R_{o(s),3}^2/R_{o(s),2}^2$ could be taken as as a probe for the three configurations. The definition of HBT correlation function and HBT radii could be found in Ref.~\cite{EPJA.56.52-JHe2020} and references therein. The hadronic rescattering time $t_{hs}$ was discussed in the evolution of fireball, which was not mentioned above. Figure~\ref{fig:radiiRatio-JHe} presented the time evolution of the ratio of the third- to the second-order HBT radii $R^2_{s(o),3}/R^2_{s(o),2}$ with two $K_T$ bins~\cite{EPJA.56.52-JHe2020}. The ratio of $R^2_{s(o),3}/R^2_{s(o),2}$ showed clearly momentum dependence and more stable with high $K_T$ which was considered due to high momentum pions likely from the center of the source at earlier stage. The key point was that the order of $R^2_{s(o),3}/R^2_{s(o),2}$ for different configurations of $^{12}$C did not change with the hadronic rescattering time $t_{hs}$, i.e. the hadronic rescattering process did not destroy the geometry properties inherited from the initial state.

\begin{figure*}[htb]
\center
	\includegraphics[angle=0,scale=0.35]{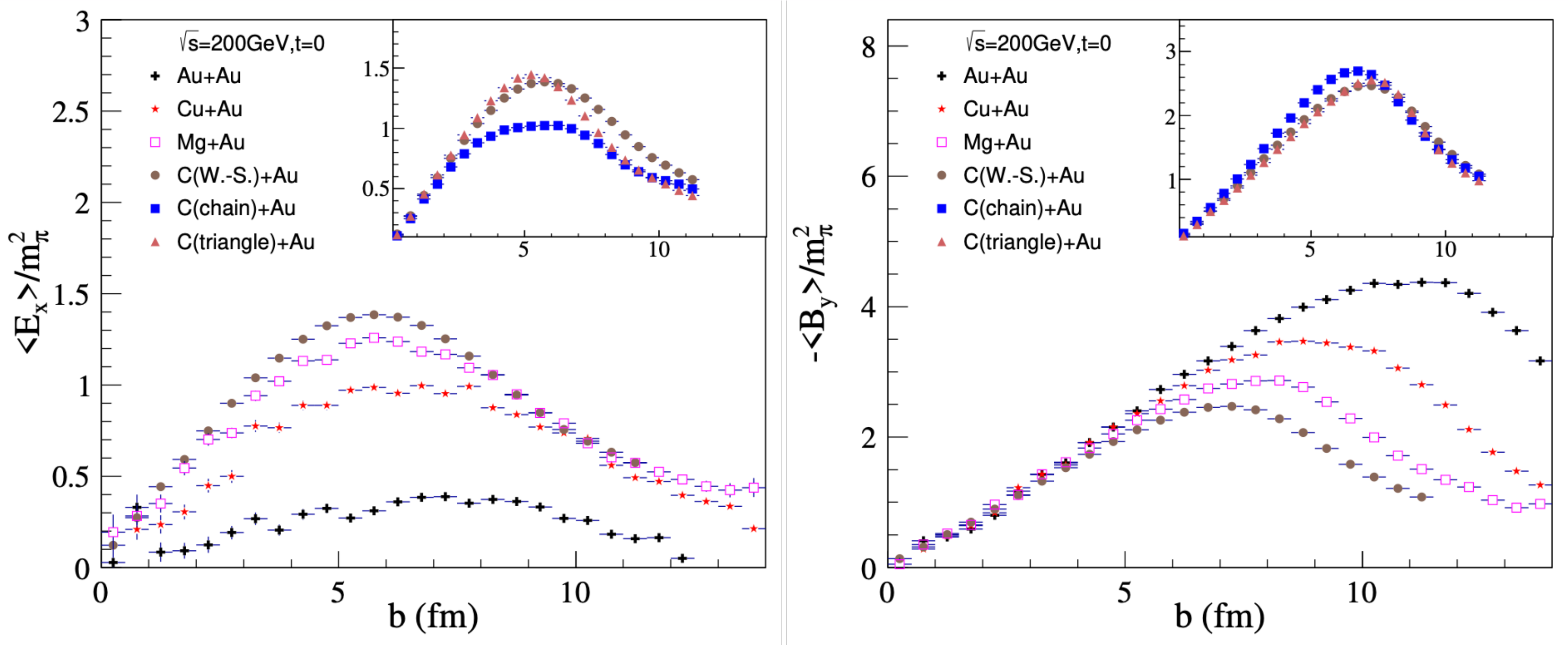}
	\caption{Impact parameter dependence of the $x$-component of electric field ($\langle E_x\rangle/m_\pi^2$) (left panel), and $y$-component of magnetic field (-$\langle B_y\rangle/m_\pi^2$) (right panel) in the collision systems Au + Au, Cu + Au, Mg + Au, and C + Au. Inset displays $\langle E_x\rangle/m_\pi^2$ (left panel) and -$\langle B_y\rangle/m_\pi^2$ (right panel) for different initial configurations of $^{12}$C~\cite{PhysRevC.99.054906-ME2019-YLCheng}.}
	\label{fig:EB_YLCheng}
\end{figure*}

It was suggested that the strong electromagnetic fields were produced in relativistic heavy-ion collisions~\cite{PhysRevLett.36.517-ME1976}. The strong electromagnetic field was considered in no-central  heavy-ion collisions and resulted in possible chiral magnetic effects~\cite{NST-28-26-CME-Hattori2017,PLB-742-296-CME-DENG2015}. If the intrinsic initial condition was taken into account, the fields would be affected by the different initial configurations of the nucleus~\cite{PhysRevC.99.054906-ME2019-YLCheng,PLB-742-296-CME-DENG2015}. The electromagnetic fields in the collision system involving $^{12}$C + $^{197}$Au, $^{24}$Mg + $^{197}$Au, $^{64}$Cu + $^{197}$Au, and $^{197}$Au + $^{197}$Au at $\sqrt{s_{NN}} = 200$ GeV~\cite{PhysRevC.99.054906-ME2019-YLCheng} were calculated by the Linard-Wiechert potential defined in ~\cite{PhysRevC.99.054906-ME2019-YLCheng},  and the $\alpha$-clustered $^{12}$C with chain and triangle structure was introduced in the calculations. Figure~\ref{fig:EB_YLCheng} showed the centrality dependence (denoted by impact parameter $b$) of the $x$-component of electric field ($\langle E_x\rangle/m_\pi^2$) (left panel), and $y$-component of magnetic field (-$\langle B_y\rangle/m_\pi^2$) (right panel) at the center of the collision zone in the mentioned systems~\cite{PhysRevC.99.054906-ME2019-YLCheng}. The inset displayed $\langle E_x\rangle/m_\pi^2$ (left panel) and -$\langle B_y\rangle/m_\pi^2$ (right panel) for different initial configurations of $^{12}$C~\cite{PhysRevC.99.054906-ME2019-YLCheng}. It indicated the chain structure of $^{12}$C affected the fields significantly. And the triangle structure of $^{12}$C reduced $\langle E_x\rangle/m_\pi^2$ in peripheral collisions rapidly and the effect could be ignored for -$\langle B_y\rangle/m_\pi^2$ by comparing with the results from the Woods-Saxon configuration. It suggested that the system scan project sheds light on hints of the exotic nuclear structure of $^{12}$C ~\cite{PhysRevC.99.054906-ME2019-YLCheng}. Of course, detailed theoretical efforts should be paid  attention to the electromagnetic effects as a probe to distinguish the initial intrinsic geometry properties.

\section{\textit{Influence of neutron skin effects}}

The geometry structure from the neutron skin effect could not be detected directly from the centrality classification but may affect the multiplicity distribution as a function of impact parameter~\cite{EPJC.77.148-2017-NS,PhysRevLett.125.222301-2020-Nch-NS,PLB-XU2021136453-NS-isobar}, dynamical process, such as electroweak processes~\cite{PLB-PAUKKUNEN201573-NS-LHC,EPJC.77.148-2017-NS,JPG.44.045104-2017-prompt-photon-NS,PhysRevC.100.024912-2019-Wpm-NS} and final collectivities~\cite{PLB-XU2021136453-NS-isobar,PhysRevC.101.061901-2020-CME-NS}.

\begin{figure*}[htb]
\center
	\includegraphics[angle=0,scale=0.9]{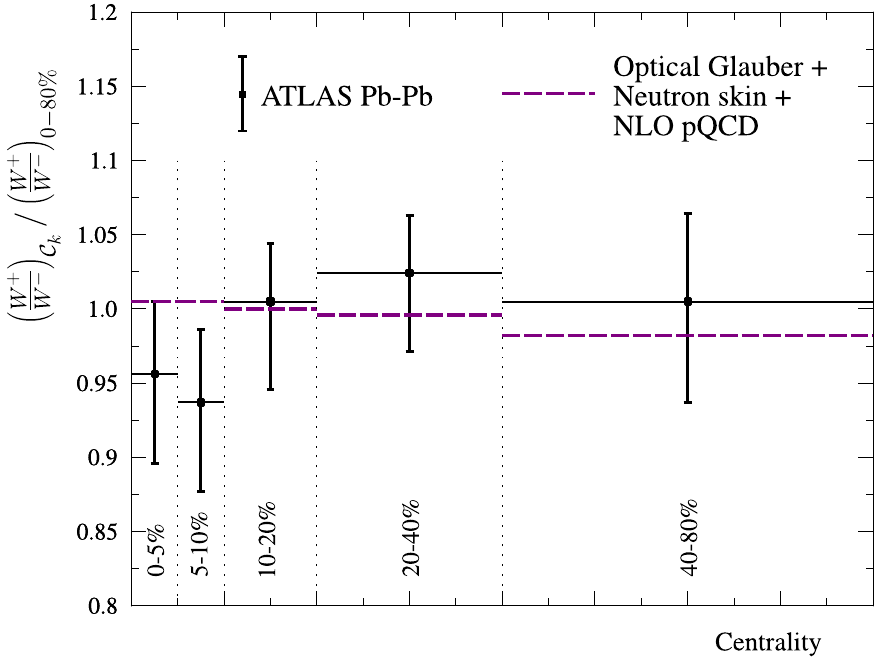}
	\caption{The centrality dependence of $W^+/W^-$ ratio as measured by the ATLAS Collaboration~\cite{ATLAS-CONF-2014-023} in comparison with the results in reference~\cite{PLB-PAUKKUNEN201573-NS-LHC}.}
	\label{fig:WpWn_Paukkunen}
\end{figure*}

The electroweak probe of the ratio of $W^{\pm}$ was proposed in Ref. ~\cite{PLB-PAUKKUNEN201573-NS-LHC} to investigate the neutron skin effect in peripheral $p$ + Pb collisions. In the Glauber modeling of the collisions, the geometry condition was considering as following for lead (Pb) nuclei: the density of nucleons in a spherical nucleus $A$ was described in 2pF distributions by equation~(\ref{eq:2pF}) and the density of protons and neutrons  differs by fixing different parameters of $a_n$, $a_p$, $d_n$ and $d_p$ if the neutron skin was taken into account. Considering the cross section in hard process $d\sigma_{nn}^{hard}$ $\neq$ $d\sigma_{pp}^{hard}$ $\neq$ $d\sigma_{pn}^{hard}$ and the production of inclusive charged leptons $l_{\pm}$ from $W^{\pm}$ $\to$ $l_{\pm}\nu$ decays (see ~\cite{PLB-PAUKKUNEN201573-NS-LHC} and references therein), the calculation reproduced the centrality dependence of $W^+/W^-$ ratio as measured by the ATLAS Collaboration~\cite{ATLAS-CONF-2014-023} as shown in figure~\ref{fig:WpWn_Paukkunen} and suggested that the $W^{\pm}$ production could be used to benchmark different experimental definitions of centrality at the LHC. Ref.~\cite{PhysRevC.100.024912-2019-Wpm-NS} extended a Monte Carlo algorithm to investigate the neutron skin effect to the ratio of $W^+/W^-$ in $p$ + Pb collisions at LHC. Except the configuration of neutron skin in Pb nucleus was taken into account, the model also introduced the color fluctuation effects and the fluctuations of the number of collisions at a given impact parameter. It was found that the effect is a factor of 2 smaller than the original estimate~\cite{PLB-PAUKKUNEN201573-NS-LHC} and suggested to extend the calculation of the $W^+/W^-$ ratio in peripheral Pb-Pb collisions as in reference~\cite{EPJC.77.148-2017-NS}.

The neutron skin effect in direct-photon was investigated in reference~\cite{JPG.44.045104-2017-prompt-photon-NS}. The total number of participant nucleons could be affected by the neutron skin thickness, especially in peripheral collisions, which was due to the nucleon-nucleon collisions, namely $pp$, $nn$ and $pn$ collision. When the photon is emitted ‘directly’ in a hard scattering, the neutron skin would contribute to the inclusive prompt photon production in nucleus-nucleus collisions where the nucleus was assumed a neutron skin nucleus. Reference~\cite{JPG.44.045104-2017-prompt-photon-NS} calculated the ‘central-to-peripheral ratio’ ($R_{cp}$) of prompt photon production with and without accounting for the neutron skin effect and it was found the neutron skin caused a characteristic enhancement of the ratio, in particular at forward rapidity in Pb+Pb collisions at LHC. The neutron-skin effect in direct-photon was also investigated in reference~\cite{SSWang} for  the reactions induced by the neutron-rich projectile $^{50}$Ca with $^{12}$C and $^{40}$Ca target in low-intermediate energy by the quantum molecular dynamics model.  The results show that more direct hard photons are produced with the increasing of the neutron skin thickness. Meanwhile, it was found 
that central-to-peripheral ratio of yield $R_{cp}$ and the rapidity dependence of multiplicity and multiplicity ratio $R_{cp}$  are apparently sensitive to neutron skin thickness. 
In reference~\cite{EPJC.77.148-2017-NS} with extending the study of reference~\cite{PLB-PAUKKUNEN201573-NS-LHC}, the results claimed that it would be demanding to unambiguously expose the neutron skin effect with direct photons~\cite{EPJC.77.148-2017-NS}. In this calculation it considered the direct photons in either in the hard process or by the fragmentation of high-$p_T$ partons from the hard process and presented the uncertainties from the parton distribution function (PDF) and 2pF in details. Instead a probe for the neutron skin was proposed by a ratio between the cross sections for negatively and positively charged high-$p_T$ hadrons~\cite{EPJC.77.148-2017-NS}.

\begin{figure*}[htb]
\center
	\includegraphics[angle=0,scale=0.9]{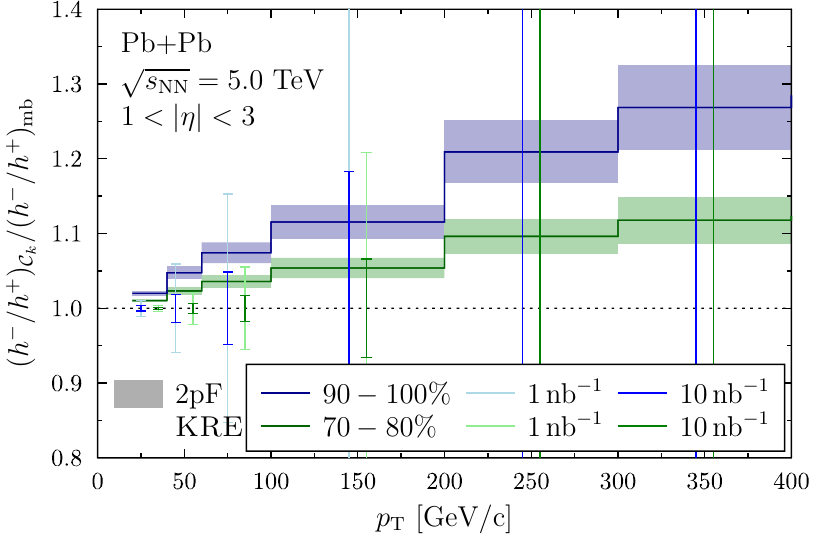}
	\caption{The expected statistical precision for 1 nb$^-1$ (light vertical bars) and for 10 $^-1$ (darker vertical bars) nucleon–nucleon luminosity in 70–80\% (green) and 90–100\% (blue) centrality classes~\cite{EPJC.77.148-2017-NS}.}
	\label{fig:Helenius-hchRatio}
\end{figure*}

The bulk properties, such as charged particle multiplicity, collective motion and chiral phenomena, were the popular observables to investigate the properties of quark-gluon plasma created in ultra-relativistic heavy-ion collisions. It became a key point in recent years if the nuclear structure affects these observables, or in other word,	whether a certain observable could be a probe to explore the exotic nuclear structure.

\begin{figure*}[htb]
\center
	\includegraphics[angle=0,scale=0.9]{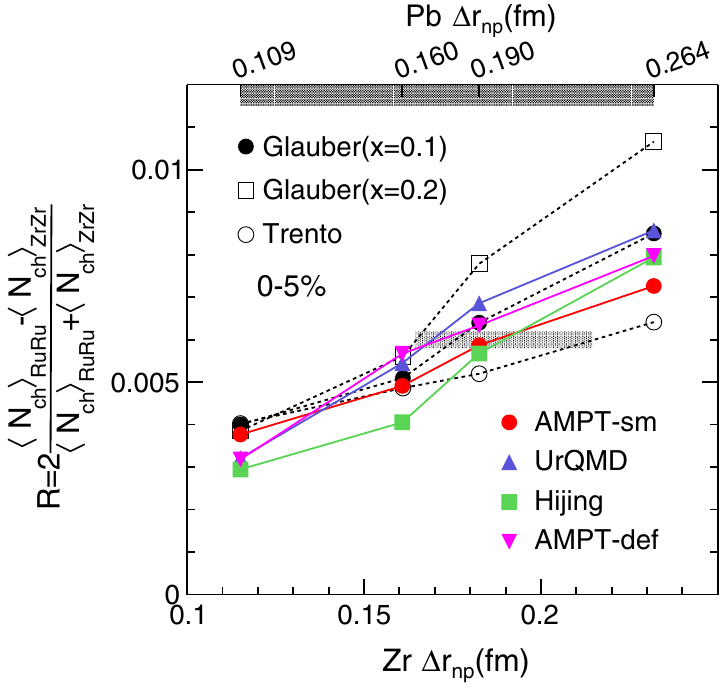}
	\caption{The relative $\langle N_{ch}\rangle$ ratio $R$ as a function of Zr neutron skin thickness~\cite{PhysRevLett.125.222301-2020-Nch-NS}.}
	\label{fig:Hanlin-Rratio}
\end{figure*}

The ratio between the cross sections for negatively and positively charged high-$p_T$ hadrons~\cite{EPJC.77.148-2017-NS}, $h^-/h^+$, was proposed to investigate the neutron skin effect in Pb+Pb collisions at LHC. With considering the uncertainties from the production mechanism and model parameters, the centrality dependence from the neutron skin effect was also clearly visible in the proposed observable~\cite{EPJC.77.148-2017-NS} as shown in figure~\ref{fig:Helenius-hchRatio} where the statistical uncertainty was also estimated as did for experiments. Recently 
Ref.~\cite{PhysRevLett.125.222301-2020-Nch-NS} developed an algorithm of the neutron skin thickness based on the symmetry energy to investigate the effect  where two different nuclear energy density functionals, namely the standard Skyrme- Hartree-Fock (SHF) model~\cite{NPA.627.710-1997-SHF-Chabanat} and the extended SHF (eSHF) model~\cite{PhysRevC.94.064326-2016-eSHF-Zhang}, were employed to describe the equation of state (EOS) and the properties of finite nuclei. Inputting the calculated neutron and proton density distributions into the transport models, AMPT-sm, AMPT-def, UrQMD~\cite{Bleicher_1999-UrQMD}, and  Hijing, separately, as the initial state, they simulated the $^{96}_{44}$Ru + $^{96}_{44}$Ru and $^{96}_{40}$Zr + $^{96}_{40}$Zr collisions at $\sqrt{s_{NN}}$ = 200 GeV~\cite{PhysRevLett.125.222301-2020-Nch-NS}. To quantify the splitting of the $N_{ch}$ tails, the relative $\langle N_{ch}\rangle$ difference between Ru + Ru and Zr + Zr collisions was defined as, $R = 2\frac{\langle N_{ch}\rangle_{RuRu}-\langle N_{ch}\rangle_{ZrZr}} {\langle N_{ch}\rangle_{RuRu}+\langle N_{ch}\rangle_{ZrZr}}$ where the tracking efficiency could be cancelled out mostly~\cite{PhysRevLett.125.222301-2020-Nch-NS}. Figure~\ref{fig:Hanlin-Rratio} showed a relatively weak model dependence of the $R$ as a function of the neutron skin thickness of Zr $\Delta r_{np}$. It suggested that measurements in the future experiment could constrain the neutron skin thickness, the symmetry energy and the parameters in EOS via comparing with the calculated results~\cite{PhysRevLett.125.222301-2020-Nch-NS}. Recently the calculation~\cite{PhysRevC.105.L011901-HL-NS-isobar} was developed to provide a method for direct measurement of the neutron skin thickness, $\Delta r_{np}$, by using net-charge multiplicities in ultra-peripheral collisions of those isobars. The calculations in references~\cite{EPJC.77.148-2017-NS,PhysRevLett.125.222301-2020-Nch-NS,PhysRevC.105.L011901-HL-NS-isobar} demonstrated that the neutron skin could affect the charged particle multiplicity distribution since the model, collision system and energy were different and all proposed to perform the  experimental measurement to reveal the the neutron skin effect in the collisions.

\begin{figure*}[htb]
\center
	\includegraphics[angle=0,scale=0.9]{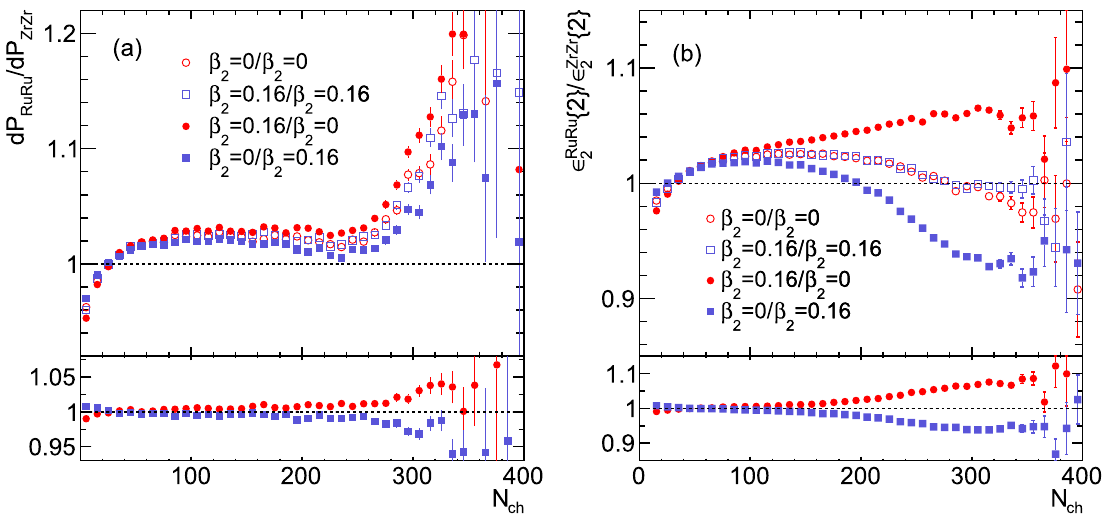}
	\caption{The Ru+Ru/Zr+Zr ratios of (a) the $N_{ch}$ distributions and (b) the $\varepsilon_2$ in relativistic isobar collisions with various combinations of $^{96}$Ru and $^{96}$Zr nuclear deformities. The lower panels show the double ratios of the deformed/spherical (and spherical/deformed) to the spherical/spherical results~\cite{PLB-XU2021136453-NS-isobar}.}
	\label{fig:kinHalo-Xu}
\end{figure*}

Ref.~\cite{PLB-XU2021136453-NS-isobar} proposed to determine the neutron skin type by $^{96}_{44}$Ru + $^{96}_{44}$Ru and $^{96}_{40}$Zr + $^{96}_{40}$Zr collisions at $\sqrt{s_{NN}}$ = 200 GeV. It was found that the Ru+Ru/Zr+Zr ratios of the $N_{ch}$ distributions and $\varepsilon_2$ in mid-central collisions were exquisitely sensitive to the neutron skin type (skin versus halo) which was clarified by the parameters in the Woods-Saxon distribution. Figure~\ref{fig:kinHalo-Xu} clearly showed the separation tail of multiplicity ratio and the eccentricity ratio due to the neutron skin type (skin vs halo) and the eccentricity ratio pattern would be reflected at the final momentum asymmetry as observables in experiment.

\section{\textit{Influence of the deformation in isobaric collisions}}

The deformation could result in the complexity of the initial geometry in ultra-relativistic heavy-ion collisions. This complexity results in the uncertainty to understand the transformation efficiency of asymmetry from the initial geometry to the final momentum space and the extraction of viscosity over entropy ratio ($\eta/s$) from the QGP created in ultra-relativistic heavy-ion collisions. In the passed two decades, amount of works raised in theory~\cite{smallSystemReview} and experiments~\cite{PhysRevLett.107.032301-2011-ALICE-vn} contributed to exploring the properties of collectivity in the fireball. Recently the  RHIC-STAR Collaboration~\cite{PhysRevC.103.064907-2021-STAR-UUFlow} reported the azimuthal anisotropy measurements of strange and multistrange hadrons in $\mathrm{U}+\mathrm{U}$ collisions at $\sqrt{{s}_{NN}}=193$ GeV and the deformation effect was studied in references~\cite{PhysRevLett.124.202301-2020-UUXeXe-th}, as well as the  ATLAS~\cite{PhysRevC.101.024906-2020-ATLAS-XeXeFlow} and the  ALICE~\cite{2021arXiv210710592A-ALICE-XeXeFlow}, respectively, reported measurements of the azimuthal anisotropy of hadrons produced in $\mathrm{Xe}+\mathrm{Xe}$ collisions at $\sqrt{{s}_{NN}}=5.44$ TeV and investigated in references~\cite{PhysRevLett.124.202301-2020-UUXeXe-th}. Correlations between transverse momentum and collective flow coefficients or among the different order of the flow coefficients were investigated thoroughly and suggested as the tools to reveal the characteristic of the initial state of the collisions~\cite{PhysRevC.103.064906-2021-vnpTCorr,PhysRevC.104.014905-2021-vnpTCorr}. All of these efforts in experiments and theory contributed strong constraints on both the initial state geometry and medium response in ultra-relativistic heavy-ion collisions. Next a brief review presents for the recently popular studies in isobaric collisions with considering the deformation in $^{96}_{44}$Ru and $^{96}_{40}$Zr.

The isobaric collisions ($^{96}_{44}$Ru+$^{96}_{44}$Ru and $^{96}_{40}$Zr+$^{96}_{40}$Zr) conducted in 2018 at the Relativistic Heavy-Ion Collider (RHIC) was motivated by the search for the chiral magnetic effect (CME) in quantum chromodynamics~\cite{CPC41.072001-2017-CME-Koch}. It was a basic problem how the initial state affects the CME signal and background from collective flow in the isobaric collision systems with similar size but different initial charge from protons. The spatial distribution of $^{96}_{44}$Ru and $^{96}_{40}$Zr in the rest frame could be written in the Woods-Saxon form as in equation~(\ref{eq:deformedWS}).

Ref.~\cite{PhysRevC.94.041901-2016-IsobaCME-Deng} calculated the magnetic field and the initial geometry eccentricity coefficient in $^{96}_{44}$Ru+$^{96}_{44}$Ru and $^{96}_{40}$Zr+$^{96}_{40}$Zr collisions at $\sqrt{s_{NN}}$ = 200 GeV by using HIJING model in which  the nuclear deformation was considered and formulated in equation~(\ref{eq:deformedWS}). The results showed more than 10\% difference in the CME signal and less than 2\% difference in the elliptic-flow-driven backgrounds for the centrality range of 20–60\%, which indicated that the isobaric collisions could provide an ideal tool to disentangle the CME signal from the background effects. 

However, Ref.~\cite{PhysRevLett.121.022301-2018-CME-NS} gave a conservative prediction of the CME signal to background by employing the density functional theory (DFT) to calculate the proton and neutron distributions as the initial spatial state. The eccentricities and magnetic fields were calculated by using the Monte Carlo–Glauber model and the DFT densities were implemented in AMPT model to simulate the elliptic flow. It was found that the results respective to the participant and reaction planes resulted in comparable differences, in the two isobaric collision systems, for eccentricities or magnetic fields and the differences are insignificant with respect to the participant plane, which possibly weakened the power of isobaric collisions for the CME search~\cite{PhysRevLett.121.022301-2018-CME-NS}.  Actually, very recently STAR Collaboration reported the first detailed results for searching for CME  by Ru+Ru to Zr+Zr isobaric collisions and found no visible CME signal \cite{STAR-PRC-2022}. Of course, this is not the end of CME story. In contrary, the data needs to more high statistics and the background needs to be more clearly understood and separated. It's a key problem to understand the background in CME which is considered mainly from the collective flow. The quadrupole and octupole deformations introduced simultaneously for the isobar systems can reproduce the centrality dependence of the $v_2$ ratio qualitatively and $v_3$ ratio quantitatively~\cite{CJZhang-arXiv210901631Z-isobar-deform}. It provides a new method to constraint or measure parameters of nuclear structure in relativistic heavy-ion collisions systems~\cite{CJZhang-arXiv210901631Z-isobar-deform}.

\begin{figure*}[htb]
\center
	\includegraphics[angle=0,scale=0.9]{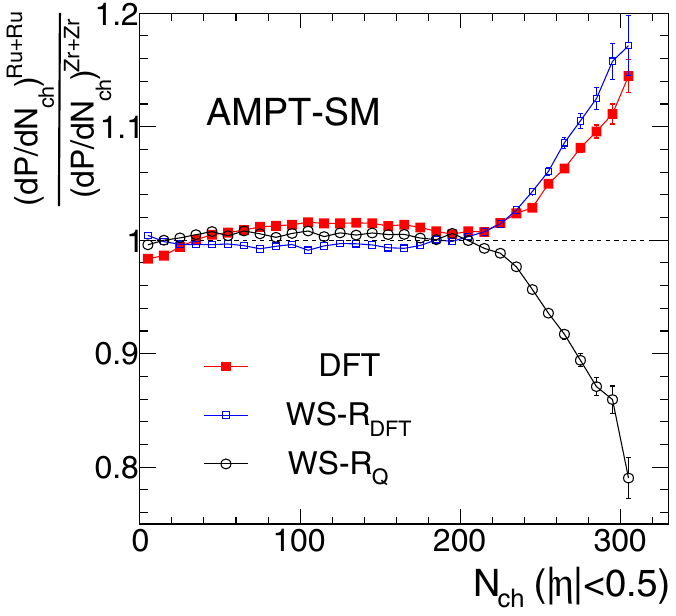}
	\caption{Ratio of the $N_{ch}$ distributions in Ru+Ru and Zr+Zr collisions simulated by AMPT-SM~\cite{PhysRevC.98.054907-2018-Nch-NS}.}
	\label{fig:hch_Hanlin}
\end{figure*}

\begin{figure*}[htb]
\center
	\includegraphics[angle=0,scale=0.9]{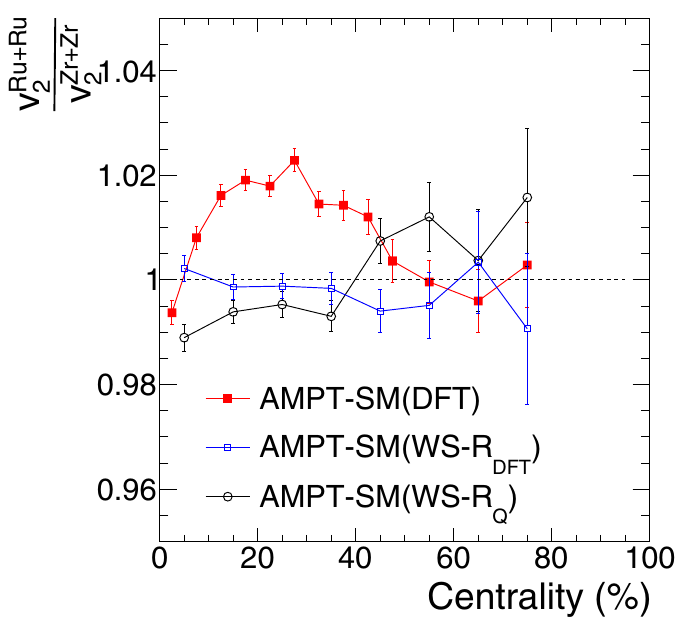}
	\caption{Ratios of charged hadron $v_2$ in Ru+Ru to Zr+Zr collisions as a function of centrality~\cite{PhysRevC.98.054907-2018-Nch-NS}.}
	\label{fig:v2RuZr_Hanlin}
\end{figure*}

Ref.~\cite{PhysRevC.99.044908-2019-IsobarFlowbkg-Schenke} developed a hydrodynamic framework which was initialized by the impact parameter-dependent Glasma model and employed UrQMD~\cite{Bleicher_1999-UrQMD} transport model as an afterburner to simulate the ultra-relativistic heavy-ion collisions, as well as isobaric collision system. The calculations estimated the background for chiral magnetic effect signal in heavy-ion collisions and predicted differences of up to 10\% on the elliptic flow of both collision systems (Ru+Ru and Zr+Zr) related to deformation.

Ref.~\cite{PhysRevC.101.061901-2020-CME-NS} introduced isospin-dependent nucleon-nucleon spatial correlations in the geometric description of both nuclei, deformation for $^{96}_{44}$Ru and the so-called neutron skin effect for the neutron-rich isobar, i.e., $^{96}_{40}$Zr in the SMASH model. The results suggested a significantly smaller CME signal to background ratio for the experimental charge separation measurement in peripheral collisions with the isobar systems than previously expected~\cite{PhysRevC.101.061901-2020-CME-NS}.

The charged hadron multiplicity distributions and collective flow were studied in the deformed isobaric collisions~\cite{PhysRevC.98.054907-2018-Nch-NS} by using AMPT model with the nuclear structure calculated by the density functional theory (DFT)~\cite{PhysRevLett.121.022301-2018-CME-NS}.
Reference~\cite{PhysRevC.98.054907-2018-Nch-NS} presented the predictions for the charged hadron multiplicity distributions in $^{96}_{44}$Ru+$^{96}_{44}$Ru and $^{96}_{40}$Zr+$^{96}_{40}$Zr collisions at RHIC. The charged hadron ratio showed in figure~\ref{fig:hch_Hanlin} and it clearly presented different trend of the centrality dependence of the ratio with nuclei radii from the DFT calculation and the mass radii~\cite{PhysRevC.94.041901-2016-IsobaCME-Deng}. The deformation effect to collective flow was presented in figure~\ref{fig:v2RuZr_Hanlin} which clearly show the elliptic flow difference for the centrality range of 20–60\%.

\section{\textit{Summary}}
A brief introduction to ultra-relativistic heavy-ion collisions and the effects from different initial state. structures were presented. The efforts in theoretical and experimental works were focused on the initial geometry asymmetry effect involving the overlap zone of the collided nuclei in collisions, the fluctuations and the intrinsic nuclear structure. In particular, impacts from $\alpha$-clustering structure, neutron skin and nuclear  deformation were discussed by hydrodynamics, transport model and some hybrid models as input in the initialization. The probes to exploring the intrinsic nuclear structure, or say how the intrinsic nuclear structure affects the final observables, were proposed by amount of theoretical works, such as collective flow, HBT correlations and charged hadron or photon production, or chiral magnetic effect in the collisions at RHIC and LHC. Although there exists model dependence or different choice of parameters, some final observables were  sensitive to the initial sate. These efforts in various theoretical survey as well as experimental observation  contributed to understand the basic problem of nuclear structure in nuclear community, and the ultra-relativistic heavy-ion collisions could be taken as a  potential platform to study the nuclear structure except the original aim of quark-gluon plasma.

\section{\textit{Acknowledgments}}
This work was supported in part by the National Natural Science Foundation of China under contract Nos. 11875066, 11890710,  11890714, 11925502, 11935001, 12147101, 11961141003, National Key R\&D Program of China under Grant No.  2018YFE0104600 and 2016YFE0100900, the Strategic Priority Research Program of CAS under Grant No. XDB34000000, the Key Research Program of Frontier Sciences of the CAS under Grant No. QYZDJ-SSW-SLH002, and the Guangdong Major Project of Basic and Applied Basic Research No. 2020B0301030008.

\bibliographystyle{apsrev4-2}
\bibliography{reference}

\end{document}